\documentclass[a4paper,11pt]{article}
\pdfoutput=1 

\usepackage{jheppub} 
\usepackage[T1]{fontenc} 
\usepackage[utf8]{inputenc}

\usepackage{xcolor}
\usepackage{upgreek}
\usepackage{bigints}
\usepackage{multirow}

\newcommand{\diag}{\ensuremath{{\rm diag}}}

\title{\boldmath Solar $\Bar{\nu}_e$ flux: Revisiting bounds on neutrino 
magnetic moments 
and solar magnetic field}

\author[a]{Evgeny Akhmedov }
\author[b,c]{and Pablo Mart\'{\i}nez-Mirav\'e }

\affiliation[a]{Max-Planck-Institut f\"ur Kernphysik, Saupfercheckweg 1, 69117 Heidelberg, Germany}
\affiliation[b]{Instituto de F{\'i}sica Corpuscular  (CSIC-Universitat de Val{\`e}ncia), Parc Cient{\'\i}fic UV, \\ \mbox{C/Catedr{\'a}tico} Jos{\'e} Beltr{\'a}n 2, Paterna 46980, Spain}
\affiliation[c]{Departament de F{\'i}sica Te\`orica, Universitat de Val{\`e}ncia, \\ C/ Dr. Moliner 50, 46100 Burjassot, Spain}

\emailAdd{akhmedov@mpi-hd.mpg.de}
\emailAdd{pamarmi@ific.uv.es}

\abstract{
The interaction of neutrino transition magnetic dipole moments with magnetic fields can give rise to the phenomenon of neutrino spin-flavour precession (SFP). For Majorana neutrinos, the combined action of SFP of solar neutrinos and flavour oscillations would manifest itself as a small, yet potentially detectable, flux of electron antineutrinos coming from the Sun. 
Non-observation of such a flux constrains the product of the neutrino magnetic moment $\mu$ and the strength of the solar 
magnetic field $B$. We derive a simple analytical expression for the expected $\bar{\nu}_e$ appearance probability in the three-flavour framework and we use it to revisit the existing experimental bounds on $\mu B$. 
A full numerical calculation has also been performed to check the validity of the analytical result. 
We also present our numerical results in energy-binned form, convenient for analyses of the data of the current and future experiments searching for the solar $\bar{\nu}_e$ flux.
In addition, we give a comprehensive compilation of other existing limits on neutrino magnetic moments and of the expressions for the probed effective magnetic moments in terms of the fundamental neutrino magnetic moments and leptonic mixing parameters.    
}

\begin{document} 
\maketitle
\flushbottom

\section{Introduction}
\label{sec:Introduction}

The study of neutrino properties is known to be a powerful tool for searching for physics beyond the Standard Model. The observation of flavour oscillations in experiments with solar, atmospheric, reactor and accelerator neutrinos imply that neutrinos have nonzero mass; this, in particular, means that they should also have magnetic dipole moments. As neutrinos are electrically neutral, they have no direct coupling to electromagnetic fields, and their electromagnetic interactions  
should arise entirely through quantum loop effects. 
In the simplest extensions of the Standard Model capable of producing nonvanishing neutrino mass, the predicted neutrino magnetic dipole moments%
\footnote{Actually, neutrinos may have magnetic and/or electric dipole moments. The former are described by the real part of the matrix of neutrino electromagnetic dipole moments $\mu$, whereas the latter, by its imaginary part. Both can cause the physical processes we consider in this paper. For brevity
we refer to $\mu$ as simply the magnetic dipole moment.
}
are too small to be probed in a foreseeable future. However, a number of models with new physics at TeV scale predict neutrino magnetic moments that may be close to the current experimental upper bounds (for a recent discussion, see e.g.\ \cite{Babu:2020ivd} and references therein).  

Photon exchange processes induced by neutrino magnetic moments can contribute to cross sections of neutrino-electron and neutrino-nucleus scattering, and in particular can affect the results of coherent elastic neutrino-nucleus scattering experiments. 
There has been an increased interest in these topics following the observation by the XENON1T experiment of an excess of low-energy electron recoil events \cite{XENON:2020rca}, which could be explained by sufficiently large neutrino magnetic moments \cite{XENON:2020rca,Miranda:2020kwy,Babu:2020ivd,Miranda:2021kre}. This excess, however, was not supported by the more recent results of XENONnT \cite{Aprile:2022vux} and is also at variance with the analysis of the results of the LUX-ZEPLIN experiment \cite{LUX-ZEPLIN:2022qhg} performed in \cite{AtzoriCorona:2022jeb}. The possibility of probing neutrino magnetic moments can be further explored in multiton xenon detectors \cite{Huang:2018nxj,Hsieh:2019hug}.

Processes induced by neutrino magnetic moments may also play an important role in astrophysical environments.  They can influence stellar evolution and can affect neutrino emission by core-collapse supernovae. Constraints on neutrino electromagnetic properties coming from non-observation of these and other processes, including constraints from the cosmic microwave background and big bang nucleosynthesis, can be found in the literature \cite{ParticleDataGroup:2020ssz}.

If neutrinos are Majorana particles, the interaction of their transition (flavour off-diagonal) magnetic moments with the solar magnetic field can result in the conversion of a fraction of left-handed $\nu_{e}$ produced in the Sun into right-handed antineutrinos $\bar{\nu}_{\mu}$ and $\bar{\nu}_{\tau}$. This spin-flavour precession (SFP) process can be resonantly enhanced by solar matter \cite{Lim:1987tk,Akhmedov:1988uk}, similarly to the resonance amplification of neutrino flavour conversion in matter (the MSW effect \cite{Wolfenstein:1977ue,Mikheyev:1985zog}). Although it is currently firmly established that the observed deficit of solar $\nu_e$ is due to the MSW effect, SFP could still be present at a subdominant level. 
The combined action of neutrino SFP and flavour oscillations would then produce a small but potentially observable flux of solar electron antineutrinos $\bar{\nu}_e$ at the Earth (see, e.g., \cite{Akhmedov:2002mf,Guzzo:2012rf} and references therein). The detection of such a flux would therefore be a clear signature of both nonzero magnetic moment and Majorana nature of neutrinos.  

Electron antineutrinos from the Sun have been searched for experimentally by KamLAND \cite{KamLAND:2003gfh,Hatakeyama:2004gm,Perevozchikov:2009bla,KamLAND:2021gvi}, Borexino \cite{Borexino:2010zht,Borexino:2019wln} and Super-Kamiokande \cite{Super-Kamiokande:2020frs} collaborations.
No excess over the expected backgrounds was found, which allowed the collaborations to establish upper bounds on the product of the transition neutrino magnetic moment and the solar magnetic field strength. In the analyses of the data presented in these papers use was made of the results of the theoretical study \cite{Akhmedov:2002mf}, which was done within a simplified 2-flavour neutrino framework and employed a standard solar model that is currently outdated. 

In the present paper we extend the theoretical analysis of \cite{Akhmedov:2002mf} to the full 3-flavour neutrino framework and also use more recent standard solar models. We develop a simple analytical approach for calculating the expected flux of $\bar{\nu}_e$ from the Sun and also solve the full system of neutrino evolution equations numerically without any simplifying approximations. Good general agreement between the results of these two approaches is found. 
All the calculations are performed for two standard solar models (low-metallicity and high-metallicity) and a number of model solar magnetic field profiles. We also study the role of various transition neutrino magnetic moments in the production of the $\bar{\nu}_e$ flux. To facilitate the extraction of the constraints on the neutrino magnetic moments and solar magnetic fields from the experimental data, we present our results in the form of simple analytical formulas as well as of ready-to-use tables of numerically calculated appearance probabilities and fluxes of solar $\bar{\nu}_e$. We then re-analyze the results of refs.~\cite{KamLAND:2003gfh,Hatakeyama:2004gm,Perevozchikov:2009bla,KamLAND:2021gvi,Borexino:2010zht,Borexino:2019wln,Super-Kamiokande:2020frs} 
using our formalism. 
For reference, we also present a compilation of bounds on neutrino magnetic moments obtained from other experiments and astrophysical observations and express the effective neutrino magnetic moments to which they are sensitive through the fundamental neutrino magnetic moments and leptonic mixing parameters.

\section{Neutrino evolution in the Sun \label{sec:InSun}}
In the absence of magnetic fields, neutrino transformations in matter are described, in the 3-neutrino framework, by the flavour evolution equation  
\begin{equation}
    i \frac{d}{dr} \nu_{fl L} = \left[U\diag(0, 2\delta, 2 \Delta)U^\dagger 
+ \diag(V_e + V_n,V_n,V_n)\right]\nu_{fl L} \,. 
\label{eq:evol1}
\end{equation}
Here $\nu_{fl L} = (\nu_{eL} \; \nu_{\mu L} \; \nu_{\tau L})^T$ is the vector of neutrino amplitudes in flavour space and $U$ is the 3-flavour leptonic mixing matrix, for which we use the standard parametrisation  
\begin{equation}
    U = O_{23}\Gamma_{\delta}O_{13}\Gamma^{\dagger}_{\delta}O_{12} \,. 
\end{equation}
Here $O_{ij}$ are the orthogonal matrices of rotation with the angle $\theta_{ij}$ in the $i-j$ plane and $\Gamma_{\delta}=\diag(1,1,e^{i\delta_{\rm CP}})$, $\delta_{\rm CP}$ being the Dirac-type CP-violating phase. In the case of Majorana neutrinos, the leptonic mixing matrix $U_M$ depends on two additional phases: $U_M = UK$, where $K = \diag(1, e^{i\lambda_{2}}, e^ {i\lambda_{3}})$. 
However, as can be seen from eq.~(\ref{eq:evol1}), these phases play no role in neutrino oscillations. We are using the notation 
\begin{equation}
    \delta = \frac{\Delta m^2_{21}}{4E}\,, \indent \indent ~~ \Delta 
= \frac{\Delta m^2_{31}}{4E},
\end{equation}
where $\Delta m_{ij}^2=m_i^2-m_j^2$ are the neutrino mass squared differences, and also denote the effective potentials due to coherent forward neutrino scattering on matter constituents by
\begin{equation}
    V_e = \sqrt{2} G_F N_e(r) \quad \text{and} \quad V_n = 
- \sqrt{2}G_F N_n(r)/2\, .
\label{eq:pot}
\end{equation}
In eq.~(\ref{eq:pot}),  $G_F$ is the Fermi constant and $N_e(r)$, $N_n(r)$ are the number densities of electrons and neutrons in matter, respectively.  

A convenient basis for considering flavour transitions in the Sun is defined through the relation
\begin{equation}
    \nu_{fl L} = O_{23}\Gamma_{\delta}O_{13}\nu_L'\,
    \label{eqn:basis}
\end{equation}
with $\nu_L' \equiv  
(\nu'_{eL} \; \nu'_{\mu L} \; \nu'_{\tau L})^T$. The neutrino evolution equation in the primed basis is then 
\begin{equation}
    i \frac{d}{dr} 
    \begin{pmatrix}
    \nu'_{eL} \\ \nu'_{\mu L} \\ \nu'_{\tau L} 
    \end{pmatrix} = \begin{pmatrix}
    2\delta s^2_{12} + c^2_{13}V_e +V_n & 2\delta s_{12}c_{12} & 
s_{13}c_{13}V_e \\ 
    2\delta s_{12}c_{12} & 2\delta c^2_{12}+ V_n & 0 \\
    s_{13}c_{13}V_e & 0 & 2\Delta + s^2_{13}V_e +V_n
    \end{pmatrix}
\!
    \begin{pmatrix}
    \nu'_{eL} \\ \nu'_{\mu L} \\ \nu'_{\tau L} 
    \end{pmatrix}\! \equiv\! H \begin{pmatrix}
    \nu'_{eL} \\ \nu'_{\mu L} \\ \nu'_{\tau L} 
    \end{pmatrix},
    \label{eq:evol2}
\end{equation}
where we used the short-hand notation $c_{ij}\equiv\cos \theta_{ij}$, $s_{ij}\equiv \sin \theta_{ij}$. 

Next, we include the effects of SFP due to the interaction of the neutrino magnetic moments with external magnetic fields. The evolution equation is then \cite{Lim:1987tk,Akhmedov:1989df,Akhmedov:1993sh}
\begin{equation}
    i \frac{d}{dr} \begin{pmatrix}
    \nu'_{L} \\ \bar{\nu}'_{ R}
    \end{pmatrix} = \begin{pmatrix}
    H & \mathcal{B} \\ \mathcal{B}^ {\dagger} & \bar{H}
    \end{pmatrix}\begin{pmatrix}
    \nu'_{L} \\ \bar{\nu}'_{R}
    \end{pmatrix}
\label{eq:evol}
\end{equation}
where $\bar{\nu}_R'=(\bar{\nu}'_{eR}\;\bar{\nu}'_{\mu R}\;
\bar{\nu}'_{\tau R})^T$ is the vector of the right-handed antineutrino amplitudes in the primed basis, $H$ is the Hamiltonian defined in the evolution equation \eqref{eq:evol2} and $\bar{H}$ is obtained from $H$ by substituting $\delta_{\rm CP} \rightarrow -\delta_{\rm CP}$, $V_e \rightarrow - V_e$ and $V_n\rightarrow-V_n$. For Majorana neutrinos, to which we restrict ourselves in the present paper, the matrix $\mathcal{B}$, describing neutrino interactions with the external magnetic field, can be written as 
\begin{equation}
    \mathcal{B} = \begin{pmatrix}
    \mathcal{B}_{e'e'} & \mathcal{B}_{e'\mu'} & \mathcal{B}_{e'\tau'}\\
    \mathcal{B}_{\mu'e'} & \mathcal{B}_{\mu'\mu'} & \mathcal{B}_{\mu' \tau'} \\
    \mathcal{B}_{\tau'e'} & \mathcal{B}_{\tau'\mu'} & \mathcal{B}_{\tau'\tau'}
    \end{pmatrix}  =\begin{pmatrix}
    0 & \mu_{e'\mu'} & \mu_{e'\tau'}\\
    -\mu_{e'\mu'} & 0 & \mu_{\mu' \tau'} \\
    - \mu_{e'\tau'} & -\mu_{\mu'\tau'} & 0
    \end{pmatrix} B_{\perp} (r) e^{i\phi(r)} 
\equiv \upmu' \cdot 
B_{\perp} (r) e^{i\phi(r)} 
    \label{eq:B} \, . 
\end{equation}
Here $\upmu'$ is the matrix of transition magnetic moments in the primed basis. To simplify notation, the matrix elements of $\upmu'$ are merely written with the primed indices as $\mu_{\alpha' \beta'}$, and the overall primes are omitted. The external magnetic field in the plane transverse to the neutrino momentum is described by the factor 
$B_{\perp}(r)e^{i\phi(r)}$, where $B_{\perp}(r)>0$ and the azimuthal angle $\phi(r)$ defines the direction of the magnetic field in this plane.%
\footnote{We ignore the contributions of the longitudinal component of the magnetic field as they  
are inversely proportional to the neutrino Lorentz factor and are thus negligible in all situations of practical interest.}

It is useful to relate the magnetic moments in the primed basis with the magnetic moments $\upmu_m$ in the neutrino mass eigenstate basis, which are of more fundamental nature: 
\begin{equation}
    \upmu'=\Gamma_\delta O_{12} K^* \upmu_{m} K^* O^T_{12} \Gamma_\delta \,.
\end{equation}
For the nonzero matrix elements of $\upmu'$ we find 
\begin{align}
    \mu_{e'\mu'} & = \mu_{12}e^ {-i\lambda_{2}}\,, 
\label{eq:emu}
\\
    \mu_{e'\tau'} & = \left(\mu_{13}c_{12} + \mu_{23}s_{12}
e^{-i \lambda_{2} }\right) e^{-i(\lambda_{3}-\delta_{\rm CP})}\,,
\label{eq:etau}\\
    \mu_{\mu' \tau'} & = \left(\mu_{23}c_{12}e^{-i\lambda_{2}} - 
\mu_{13}s_{12}\right)e^{-i(\lambda_{3}-\delta_{\rm CP})} \, .
\label{eq:mutau}
\end{align}

The evolution equation (\ref{eq:evol}) can now be written in more detailed 
form as  
\begin{subequations}
\begin{align}
    i\frac{d}{dr}\nu'_{eL} &= H_{e'e'}\nu'_{eL} + H_{e'\mu'}\nu'_{\mu L} + 
    H_{e'\tau'}\nu'_{\tau L} + \mathcal{B}_{e'\mu'}\bar{\nu}'_{\mu R} + 
    \mathcal{B}_{e'\tau'}\bar{\nu}'_{\tau R} \, ,
    \\
    i\frac{d}{dr}\nu'_{\mu L} &= H_{\mu'e'}\nu'_{eL} + H_{\mu'\mu'}
    \nu'_{\mu L}  + \mathcal{B}_{\mu'e'}\bar{\nu}'_{e R} + 
    \mathcal{B}_{\mu'\tau'}\bar{\nu}'_{\tau R} \, ,
    \\
    i\frac{d}{dr}\nu'_{\tau L} &= H_{\tau'e'}\nu'_{eL} + H_{\tau'\tau'}
    \nu'_{\tau' L} + \mathcal{B}_{\tau'e'}\bar{\nu}'_{eR} + 
    \mathcal{B}_{\tau'\mu'}\bar{\nu}'_{\mu R}\, , 
    \\
    i\frac{d}{dr}\bar{\nu}'_{e R} &= \bar{H}_{e'e'}\bar{\nu}'_{eR} + 
    \bar{H}_{e'\mu'}\bar{\nu}'_{\mu R} +\bar{H}_{e'\tau'}\bar{\nu}'_{\tau R}  
    - \mathcal{B}^*_{e'\mu'} \nu'_{\mu L} -\mathcal{B}^*_{e'\tau '}
    \nu'_{\tau L}\, , 
    \\  
    i\frac{d}{dr}\bar{\nu}'_{\mu R} &= \bar{H}_{\mu'e'}\bar{\nu}'_{eR} + 
    \bar{H}_{\mu'\mu'}\bar{\nu}'_{\mu R}  - \mathcal{B}^*_{\mu'e'} \nu'_{e L} 
    - \mathcal{B}^*_{\mu'\tau '}\nu'_{\tau L}\, ,
    \\
    i\frac{d}{dr}\bar{\nu}'_{\tau R} &= \bar{H}_{\tau'e'}\bar{\nu}'_{eR} 
    +\bar{H}_{\tau'\tau'}\bar{\nu}'_{\tau R} -\mathcal{B}^*_{\tau'e'} 
    \nu'_{e L} - \mathcal{B}^*_{\tau'\mu '}\nu'_{\mu L} \, . 
\end{align}
\label{eq:evol3}
\end{subequations}
Here we have taken into account that the diagonal elements of the matrix $\mathcal{B}$ vanish and also that $H_{\mu'\tau'}=H_{\tau'\mu'} =\bar{H}_{\mu'\tau'} = \bar{H}_{\tau'\mu'}=0$. 

\subsection{\label{sec:Approx} Approximate analytical solution of the evolution equation}
Because the diagonal magnetic moments of Majorana neutrinos vanish, direct conversion of the left-handed electron neutrinos produced in the Sun into $\bar{\nu}_{eR}$ is not possible. Still, $\nu_{eL}\to\bar{\nu}_{eR}$ transitions can proceed via two-step processes,  
\begin{subequations}
\begin{align}
    \nu_{eL} \overset{\rm osc.}{\longrightarrow} \nu_{\mu L} 
\overset{\rm SFP}{\longrightarrow} \bar{\nu}_{eR}\,,
\label{eq:trans1}
\\
    \nu_{eL} \overset{\rm SFP}{\longrightarrow} \bar{\nu}_{\mu R} 
    \overset{\rm osc.}{\longrightarrow} \bar{\nu}_{eR} \, ,
\label{eq:trans2}
\end{align}
\end{subequations}
and similarly for transitions through the $\nu_{\tau L}$ and 
$\bar\nu_{\tau R}$ intermediate states. However, inside the Sun such conversions should be heavily suppressed because the amplitudes of the processes (\ref{eq:trans1}) and (\ref{eq:trans2}) are of opposite sign and 
nearly cancel each other \cite{Akhmedov:1989df,Akhmedov:1993sh}. 
For the same reasons, the transitions $\nu_{eL}'\to\bar{\nu}_{eR}'$ between the primed states are also suppressed. 

The transitions $\nu_{eL}\to \bar{\nu}_{eR}$ through the processes (\ref{eq:trans1}) and (\ref{eq:trans2}) (and similar transitions with $\nu_{\tau L}$ and $\bar\nu_{\tau R}$ intermediate states) will, however, not be suppressed if the flavour conversions and SFP occur in spatially separated regions. 
Because magnetic fields outside the Sun are very weak, we are left with the possibility of the transition chain (\ref{eq:trans2}), with SFP taking place inside the Sun and the subsequent flavour conversions occurring on the flight between the Sun and the Earth. To calculate the flux of solar $\bar{\nu}_{eR}$ on the Earth we therefore first need to find the fluxes of $\bar{\nu}_{\mu R}'$ and $\bar{\nu}_{\tau R}'$ at the surface of the Sun.
 
We shall now develop an approximate analytical approach to this problem. First, basing on the above arguments, we neglect $\nu_{eL}'\to \bar{\nu}_{eR}'$ conversions inside the Sun.
We therefore omit the evolution equation for $\bar{\nu}_{eR}'$ as well as any terms containing the $\bar{\nu}_{eR}'$ amplitude from the equation system (\ref{eq:evol3}). 
Next, we neglect the terms containing $H_{e'\tau'}=H_{\tau' e'}$ (and $\bar{H}_{e'\tau'} = \bar{H}_{\tau'e'})$, since they are much smaller than the diagonal elements $H_{\tau' \tau'}$ and 
$\bar{H}_{\tau'\tau'}$, which means that flavour transitions, caused by the above-mentioned off-diagonal terms, are strongly suppressed. 
Finally, we take into account that the effects of SFP of solar neutrinos are expected to be small and restrict ourselves to leading order in perturbation theory in $\mathcal{B}$. As we are interested in calculating the amplitudes $\bar{\nu}_{\mu R}'$ and $\bar{\nu}_{\tau R}'$, whose evolution equations 
contain the amplitudes $\nu_{eL}'$, $\nu_{\mu L}'$, and $\nu_{\tau L}'$ multiplied by the elements of the $\mathcal{B}$ 
matrix, the amplitudes of these left-handed states should be found to zeroth order in $\mathcal{B}$. Applying these approximations to eq.~(\ref{eq:evol3}), we find the simplified evolution equations
\begin{subequations}
\begin{align}
    i\frac{d}{dr}\nu'_{eL} &= \left(2\delta s^2_{12} + c^2_{13}V_e +V_n\right) 
\nu'_{eL} + 2\delta s_{12}c_{12}\nu'_{\mu L}\,,
    \label{eqn:nuel}
\\
    i\frac{d}{dr}\nu'_{\mu L} &= 2\delta s_{12}c_{12} \nu_{e'L} + 
\left(2\delta c^2_{12}+V_n\right)\nu'_{\mu L}\,,
    \label{eqn:numul}
\\
    i\frac{d}{dr}\nu'_{\tau L} &= \left(2\Delta + s^2_{13}V_e + V_n\right)
    \nu'_{\tau  L}\,,
    \label{eqn:nutaul}
\\
    i\frac{d}{dr}\bar{\nu}'_{\mu R} &= \left(2\delta c^2_{12} -V_n\right)
    \bar{\nu}'_{\mu R} + \mu^*_{e'\bar{\mu}'} B_{\perp}e^{-i\phi}\nu'_{eL} - 
    \mu^*_{\mu' \bar{\tau}'}B_{\perp}e^{-i\phi}\nu'_{\tau L}\,,
    \label{eqn:numur}
\\
    i\frac{d}{dr}\bar{\nu}'_{\tau R} &= \left(2\Delta -s^2_{13}V_e - 
    V_n\right)\bar{\nu}'_{\tau R} + \mu^ *_{e'\bar{\tau}'}B_{\perp}e^{-i\phi}
    \nu'_{eL} + \mu^*_{\mu' \bar{\tau}'}B_{\perp}e^{-i\phi}\nu'_{\mu L}\,.
    \label{eqn:nutaur}
\end{align}
\end{subequations}
We first note that the first two of these equations, describing the evolution of the amplitudes $\nu_{eL}'$ and $\nu_{\mu L}'$, decouple from the rest of the system and can be solved independently. This essentially reduces to solving the MSW problem for solar neutrinos. We therefore employ the adiabatic approximation, which is known to work very well in this case, and obtain
\begin{equation}
    \nu'_{eL}(r) = c_{13}\left[\cos \Tilde{\theta}(r_0)\cos \Tilde{\theta}(r) 
    e^{-i\int_{r_0}^ r E_1 dr'} + \sin \Tilde{\theta}(r_0)\sin
    \Tilde{\theta}(r) e^{-i\int_{r_0}^ r E_2 dr'}\right], ~~~
\end{equation}
\begin{equation}
    \nu'_{\mu L}(r) = c_{13}\left[-\cos \Tilde{\theta}(r_0)\sin 
\Tilde{\theta}(r) e^{-i\int_{r_0}^ r E_1 dr'} + \sin \Tilde{\theta}(r_0)
\cos \Tilde{\theta}(r) e^{-i\int_{r_0}^ r E_2 dr'}\right] \,.
\end{equation}
Here $r_0$ is the coordinate of the neutrino production point, 
$\Tilde{\theta}(r)$ is the effective mixing angle in matter which can be found from the relation 
\begin{equation}
    \cos 2\Tilde{\theta}(r)=\frac{\cos2\theta_{12} - c^2_{13}V_e/2\delta}
{\sqrt{\left(\cos2\theta_{12}- c^2_{13}V_e/2\delta\right)^ 2 
+ \sin^2 2\theta_{12}}}\,,
\label{eq:tildetheta}
\end{equation}
and we have defined 
\begin{equation}
    E_{1,2} \equiv \delta + c^2_{13}\frac{V_e}{2} +V_n \mp \sqrt{\left(\delta
\cos2\theta_{12}- c^2_{13}V_e/2\right)^ 2 + \delta^ 2\sin^2 2\theta_{12}}\,.
\label{eq:E12}
\end{equation}
Note that the initial conditions $\nu_{eL}(r_0)=1$, $\nu_{\mu L}(r_0)=\nu_{\tau L}(r_0)=0$ translate, in the primed basis, to $\nu'_{eL}(r_0) = c_{13}$, $\nu'_{\mu L} (r_0) = 0$, $\nu'_{\tau L} (r_0) = s_{13}$.  

The evolution equation for the amplitude $\nu_{\tau L}'$ completely decouples from the rest of the system and its solution is 
\begin{equation}
    \nu'_{\tau L}(r) = s_{13} e^{-i\int_{r_0}^ r\left(2\Delta + s^2_{13}V_e 
+ V_n\right)dr'}.
\end{equation}

Now that $\nu_{eL}'$, $\nu_{\mu L}'$ and $\nu_{\tau L}'$ are found, it straightforward to solve eqs.~(\ref{eqn:numur}) and (\ref{eqn:nutaur}). For the values of the amplitudes $\bar{\nu}'_{\mu R}$ and $\bar{\nu}'_{\tau R}$ at the surface of the Sun we obtain   
\begin{subequations}
\begin{align}
    \bar{\nu}'_{\mu  R} (R_{\odot}) =& \int_{r_0}^ {R_\odot} B_{\perp} (r) 
     \left[c_{13}\mu^ *_{e' \mu'} \cos \Tilde{\theta} (r_0) \cos 
     \Tilde{\theta} (r) e ^ {-i g_1 (r)} \right.\,, \nonumber \\ 
   &+ c_{13}\mu^ *_{e' \mu'} \sin \Tilde{\theta} (r_0) \sin \Tilde{\theta}(r) 
e^{-i g_2 (r)}  \left. - s_{13}\mu ^*_{\mu' \tau'} e ^ {-ig_3(r)} \right] dr\,,
\label{eq:numur1}
\\
    \bar{\nu}'_{\tau R} (R_{\odot}) =& \int_{r_0}^ {R_\odot} B_{\perp} (r) 
c_{13} \cos \Tilde{\theta} (r_0) \left(\mu^*_{e'\tau'}\cos \Tilde{\theta} (r) 
-\mu^*_{\mu'\tau'}\sin \Tilde{\theta}(r)\right) e^{-ig_4(r)} dr\,, \nonumber \\ 
    &+ \int_{r_0}^ {R_\odot} B_{\perp} (r) c_{13} \sin\Tilde{\theta} (r_0) 
\left(\mu^*_{e'\tau'}\sin \Tilde{\theta} (r) + \mu^*_{\mu'\tau'}
\cos \Tilde{\theta} (r)\right) e^{-ig_5(r)} dr\,,
\label{eq:nutaur1}
\end{align}
\label{eqn:amplitud}
\end{subequations}
where we have defined 
\begin{subequations}
\begin{align}
    g_{1,2}(r) & \equiv
 \phi + \int_{r_0} ^ r \left[E_{1,2} - \left( 2c^2_{12}
\delta - V_n \right)\right]dr'\,,
\label{eq:g1}
\\
    g_{3}(r) &\equiv \phi + \int_{r_0} ^ r \left[\left(2\Delta +s^2_{13}V_e 
+V_n \right) - \left( 2c^2_{12}\delta - V_n \right) \right]dr'\,,
\label{eq:g2}
\\
    g_{4,5}(r) &\equiv \phi + \int_{r_0} ^ r \left[ E_{1,2}-\left(2\Delta 
-s^2_{13}V_e -V_n \right) \right] dr'\,,
\label{eq:g3}
\end{align}
\end{subequations}
and have dropped the irrelevant overall phase factors from the expressions for $\bar{\nu}'_\mu$ and $\bar{\nu}'_\tau$. 
Such inconsequential phase factors will also be consistently 
omitted in what follows. 
 
\subsubsection{Analytical expressions for the amplitudes}
\label{sec:Amplitude} 
The integrals in eq.~\eqref{eqn:amplitud} are of general form
\begin{equation}
    I = \int_{a}^b f(x) e^{-ig(x)}dx\,,
\label{eq:int1}
\end{equation}
where $f(x)$ is a slowly varying function of coordinate and 
$|g'(x)|$ is large except possibly in the vicinity of a finite number of points in the interval $(a,b)$. Such integrals get their main contributions from the endpoints of the integration intervals and from the stationary phase points where $g'(x)=0$, if any \cite{Erdelyi} (see also section~\ref{sec:non-tw} below).  Let us first check if stationary phase points for the integrals in eqs.~(\ref{eq:numur1}) and (\ref{eq:nutaur1}) exist. 

The evolution equation (\ref{eq:numur1}) for the amplitude $\bar{\nu}_{\mu R}'$ depends on the phases $g_1$, $g_2$ and $g_3$. The stationary phase conditions are $\frac{d}{dr}g_{1,2} = 0$ and $\frac{d}{dr}g_{3} = 0$ or, respectively, 
\begin{equation}
\frac{d\phi}{dr} =
2c^2_{12}\delta -V_n -E_{1,2}\,,\qquad\quad~~
\label{eq:statpoint1}
\end{equation}
\begin{equation}
\frac{d\phi}{dr} =
2c^2_{12}\delta -2V_n -2\Delta -s^2_{13}V_e\,. 
\label{eq:statpoint2}
\end{equation}
The stationary phase conditions for the integrals in eq.~(\ref{eq:nutaur1}) are 
$\frac{d}{dr}g_{4,5} = 0$, or  
\begin{equation}
\frac{d\phi}{dr} = 2\Delta - s^2_{13}V_e -V_n -E_{1,2} = 0\,.
\label{eq:statpoint3}
\end{equation}
Consider first eq.~(\ref{eq:statpoint3}). The term $2\Delta$ on its right hand side is at least an order of magnitude larger than the other terms (note that $2\Delta \sim 10^{-9} - 10^ {-10}$ eV). For the solution of this equation to exist, $d\phi/dr$ should be of the same order of magnitude, which corresponds to $\sim 1-10$ rad/km. While short-scale stochastic magnetic fields in the Sun may possibly have such rapid twists, it is unlikely that this is possible for large-scale fields relevant for SFP. 
This means that no stationary phase points 
are expected for the integrals in eq.~(\ref{eq:nutaur1}), and they should receive their main contributions from the endpoints of the integration interval. The same arguments apply to eq.~(\ref{eq:statpoint2}) and the integral containing $g_3$ in eq.~(\ref{eq:numur1}). 

Let us now examine eq.~(\ref{eq:statpoint1}). Using eq.~(\ref{eq:E12}), one can reduce it to  
\begin{equation}
    d\phi/dr +2V_n+c^2_{13}V_e - 2\delta \cos2\theta_{12} = \frac{\delta^2 
\sin ^2 2\theta_{12}}{d\phi/dr + 2 V_n} \,, 
\end{equation}
which has a solution as long as
\begin{equation}
    1 + \sin^2 2\theta_{12} \frac{c^2_{13}V_e}{d\phi/dr +2V_n} \geq 0 \, .
\label{eq:twistcond1}
\end{equation}
This condition is satisfied if
\begin{align}
    \frac{d \phi}{dr} >  2|V_n| \quad \text{or} \quad
    -\frac{d \phi}{dr} \geq  c^ 2_{13}V_e 
\left(\sin^2 2\theta_{12}- \frac{1- Y_e}{c^ 2_{13}Y_e}\right), 
\label{eq:twistcond2}
\end{align}
where $Y_e$ is the number of electrons per nucleon in the medium. As $Y_e$ varies between 0.67 and 0.88 in the Sun \cite{Serenelli:2009yc}, it is easy to see that the expression in the brackets in the second condition in (\ref{eq:twistcond2}) is positive and on the order of 0.3 -- 0.7; therefore, for non-twisting magnetic fields the stationary phase condition cannot be fulfilled. In fact, it requires $|d\phi/dr|$ to be of the same order of magnitude as $V_e$ and $|V_n|$, which vary from $\sim 7\times 10 ^{-12} \text{ eV}$ near the neutrino production point to zero at the surface of the Sun, where the solar magnetic field nearly vanishes as well.  
One can see that the stationary phase condition can be fulfilled, for instance, for magnetic fields with constant twist $|d\phi/dr| \sim 10/R_\odot \sim 3\times 10^ {-15} \text{ eV}$ \cite{Akhmedov:1993sh}.

We will first focus on the case in which the magnetic fields in the Sun are either non-twisting or they twist slowly enough, so that no stationary phase points exist. 
Effects of possible existence of stationary phase points in the scenario with fast twisting magnetic fields will be discussed in section~\ref{sec:twist}. 

\subsubsection{\label{sec:non-tw}Non-twisting or slowly twisting magnetic fields} 
In this case, the integrals in eqs.~(\ref{eq:numur1}) and (\ref{eq:nutaur1}) are dominated by the contributions from the endpoints of the integration intervals. To evaluate such contributions to an integral of the type (\ref{eq:int1}), we integrate it by parts. Integrating two times one finds  
\begin{align}
    \int_{a}^ {b}f(x) e^ {-ig(x)} dx = \left[\left(i \frac{f(x)}{g'(x)} + 
\frac{f'(x)}{g'(x)^2} - \frac{f(x)g''(x)}{g'(x)^3}\right)e^{-i g(x)}\right]_a ^ b + \mathcal{O}\left(\frac{1}{g'(x)^3}\right)\, .
\label{eq:appr1}
\end{align}
It follows from the definitions of the phases $g$ in eqs.~(\ref{eq:g1})-(\ref{eq:g3}) and eqs.~(\ref{eq:pot}) and~(\ref{eq:E12}) that in the case of interest to us the condition 
\begin{equation}
|g''(x)|^2/g'(x)^2 \ll 1
\label{eq:cond1}
\end{equation}
is satisfied for all $g_i(x)$ ($i=1,...,5)$. Therefore, the third term in the brackets in eq.~(\ref{eq:appr1}) can be neglected compared to the first term.   
In addition, the first term dominates over the second one provided that 
\begin{equation}
    \Bigg|\frac{f(x)}{f'(x)}\Bigg| \gg \frac{1}{|g'(x)|}.
\label{eq:cond2}
\end{equation}
Consider the left hand side of this inequality. It is essentially the scale height of the function $f$, i.e.\ the characteristic distance over which it varies significantly. 
As follows from (\ref{eq:numur1}) and (\ref{eq:nutaur1}), in the case under discussion $f(r)\propto B_\perp(r)$ times $\sin\tilde{\theta}(r)$ or $\cos\tilde{\theta}(r)$. Because the effective mixing angle $\tilde{\theta}$ is a slowly varying function of coordinate inside the Sun (which actually justifies using the adiabatic approximation for flavour conversions),  the scale height of $f(r)$ essentially coincides with the scale height of the solar magnetic field, $L_B\equiv B_\perp(r)/|B_\perp'(r)|$.    
Therefore eq.~(\ref{eq:cond2}) reduces to 
\begin{equation}
    L_B 
\gg \frac{1}{|g'_i(r)|} \, 
\end{equation}
for each of the five $g_i(r)$ defined. For the propagation of neutrinos in the Sun, these conditions are satisfied if $L_B \gg 10^ {-4} R_\odot$. 
Magnetic fields with scale heights as small as $L_B\lesssim 10^{-4}R_\odot$ can only exist over very short distances in the Sun, and so they cannot lead to any sizeable SFP. We therefore only consider large-scale solar magnetic fields, which satisfy (\ref{eq:cond2}). 
As a consequence, it is justified to retain only the first term in the brackets in eq.~(\ref{eq:appr1}). In addition, we take into account that the magnetic field strength at the surface of the Sun $B_\perp(R_\odot)$ is very weak and consider only the contribution of the neutrino production point $r=r_0$. Eqs.~(\ref{eq:numur1}) and (\ref{eq:nutaur1}) thus yield 
\begin{subequations}
\begin{align}
    \bar{\nu}'_{\mu  R} (R_\odot) \simeq & \, B_\perp (r_0) \left[c_{13}
\mu^*_{e'\mu'}\left(\frac{\cos^2\tilde{\theta}(r_0)}{g'_1(r_0)} + 
\frac{\sin^2\tilde{\theta}(r_0)}{g'_2(r_0)}\right) - 
\frac{s_{13}\mu^*_{\mu' \tau'}}{2\Delta}\right],
\label{eq:numur2}
\\
    \bar{\nu}'_{\tau R}(R_\odot) \simeq & \, B_\perp (r_0)\frac{c_{13}
\mu^*_{e' \tau'}}{2\Delta} \, ,
\label{eq:nutaur2}
\end{align}
\end{subequations}
where we have taken into account that $g'_3 \sim 2\Delta$, $g'_{4,5} 
\sim - 2\Delta$ and that
\begin{equation}
   \bigg|\frac{1}{g'_4}-\frac{1}{g'_5} \bigg|  \ll \bigg| \frac{1}{g'_{4,5}} 
\bigg|.
\end{equation}
Notice that setting $\theta_{13} = 0$ and neglecting $\cos\tilde{\theta}(r_0)$ compared with $\sin\tilde{\theta}(r_0)$ one recovers the expression for the amplitude of $\bar{\nu}'_{\mu R}$ found in \cite{Akhmedov:2002mf}.

\subsubsection{\label{sec:twist}Fast-twisting magnetic fields} 
Consider now the case when one of the conditions in \eqref{eq:twistcond2} is satisfied, which requires the solar magnetic field to be sufficiently fast twisting. 
Let $r_1$ and $r_2$ be such that $g'_1(r_1) = 0$ and $g'_2(r_2) =0$. The contribution of these stationary phase points to the amplitude $\bar{\nu}_{\mu R}'(R_\odot)$ in eq.~(\ref{eq:numur1}) is   
\begin{align}
    \bar{\nu}'_{\mu R}(R_\odot) =&\, c_{13}\mu^*_{e'\mu'}
\left[\cos \tilde{\theta} (r_0) \cos \tilde{\theta} (r_1)B_\perp(r_1)
\sqrt{\frac{2\pi}{g''_1(r_1)}}\right. \nonumber \\ 
& \left.+ \sin \tilde{\theta} (r_0) \sin \tilde{\theta} (r_2)B_\perp(r_2)
\sqrt{\frac{2\pi}{g''_2(r_2)}}e^ {-i (g_2(r_2) -g_1(r_1))}\right]. 
\end{align}
{}From eq.~(\ref{eq:cond1}) it follows that it strongly dominates over the contributions (\ref{eq:numur2}) that come from the endpoints of the integration interval, which can therefore be neglected in this case.    
As there are no stationary phase point contributions to the amplitude $\bar{\nu}_{\tau R}'$, it is still given by eq.~(\ref{eq:nutaur2}), just as in the cases of non-twisting or slow-twisting magnetic fields. 
Note that the validity of this approximation still relies on the assumption that there is no $\nu_e \rightarrow \bar{\nu}_e$ transitions in the Sun. In the presence of fast-twisting magnetic fields, this might not be accurate enough \cite{Akhmedov:1993sh}. 

From now on we will constrain ourselves to the case of non-twisting or slowly twisting magnetic fields, in which there are no stationary phase point contributions to the amplitudes $\bar{\nu}_{\mu R}'$  and $\bar{\nu}_{\tau R}'$.   As follows from the above discussion, this may only reduce these amplitudes, and therefore will make our upper bound on $\mu B_\perp$ more conservative.

\subsubsection{Solar electron antineutrino flux on the Earth}
\label{sec:Flux}
Once the amplitudes $ \bar{\nu}_{\mu R}'$ and $\bar{\nu}_{\tau R}'$ on the surface of the Sun have been calculated, one can compute the expected flux of $\bar{\nu}_{e R}$ that reaches the Earth. 
As the magnetic field in the space between the Sun and the Earth is negligible, neutrino evolution en route to the Earth reduces to pure flavour transformations.  Due to coherence loss, solar neutrinos (or antineutrinos) arrive at the Earth as incoherent sums of mass eigenstates \cite{Dighe:1999id}.
The probability that a $\nu_{eL}$ produced in the Sun will reach the Earth as $\bar{\nu}_{eR}$ is therefore 
\begin{equation}
  P(\nu_{eL} \rightarrow \bar{\nu}_{eR}) = |U_{e1}|^ 2 |\bar{\nu}_{1\oplus}|^2 
+|U_{e2}|^ 2 |\bar{\nu}_{2\oplus}|^ 2 + |U_{e3}|^ 2 |\bar{\nu}_{3\oplus}|^2\,,
\label{eq:P1}
\end{equation}
where $\bar{\nu}_{i\oplus}$ ($i=1,2,3$) are the amplitudes of the antineutrino mass eigenstates reaching the Earth.
These amplitudes are related to those in the primed basis by 
$\bar{\nu}'_R = \Tilde{U} \bar{\nu}_R$ with  
$\Tilde{U} = \Gamma^{\dagger}_\delta O_{12}$, where  
$\bar{\nu}_R' = (\bar{\nu}'_{eR}, \; \bar{\nu}'_{\mu R}, \; 
\bar{\nu}'_{\tau R})^T$ and $\bar{\nu}_R = (\bar{\nu}_{1R}, \; 
\bar{\nu}_{2R}, \; \bar{\nu}_{3R} )^T$. 
Therefore  
\begin{equation}
   |\bar{\nu}_{i\oplus}|^ 2 = |\Tilde{U}_{\mu ' i}|^ 2 |\bar{\nu}'_{\mu  R }
(R_\odot)|^ 2 + |\Tilde{U}_{\tau ' i}|^2 |\bar{\nu}'_{\tau  R} (R_\odot)|^2 \,,
\end{equation}
and eq.~(\ref{eq:P1}) for the electron antineutrino appearance probability can be rewritten as 
\begin{align}
  P (\nu_{eL} \rightarrow \bar{\nu}_{eR}) = \frac{1}{2}c^2_{13}\sin^2 
2\theta_{12} |\bar{\nu}_{\mu R}'(R_\odot)|^ 2 +s^2_{13}|\bar{\nu}_{\tau R}' 
(R_\odot)|^2.
\end{align} 
Substituting here the approximate analytical expressions for the amplitudes $\bar{\nu}_{\mu R}'$ and $\bar{\nu}_{\tau R}'$ from eqs.~(\ref{eq:numur2}) and (\ref{eq:nutaur2}), we find 
\begin{align}
P (\nu_{eL} \rightarrow \bar{\nu}_{eR})  = \frac{1}{2} c^2_{13}\sin^2 
2\theta_{12}B^2_\perp (r_0)\left[c^2_{13}|\mu_{e'\mu'}|^2 
\Bigg(\frac{\cos^2\tilde{\theta}(r_0)}{g'_1(r_0)} + \frac{\sin^2\tilde{\theta}
(r_0)}{g'_2(r_0)}\Bigg) ^ 2  \right. \nonumber \\  \left. + 
\Bigg( \frac{s_{13} |\mu_{\mu' \tau'}|}{2\Delta}\Bigg)^ 2 -2{\rm Re} 
\{ \mu_{e'\mu'}^* \mu_{\mu' \tau'} \}
\left(\frac{\cos^2\tilde{\theta}(r_0)}{g'_1(r_0)} + \frac{\sin^2\tilde{\theta}
(r_0)}{g'_2(r_0)}\right) \frac{s_{13} c_{13}}{2\Delta}  \right] \nonumber \\ 
+ s^2_{13}B^2_\perp (r_0)\Bigg( \frac{c_{13} |\mu_{e'\tau'}|}
{2\Delta}\Bigg)^ 2\, ,
    \label{eqn:peebar}
\end{align}
where the terms containing $|\mu_{\mu'\tau'}|^2$ and $|\mu_{e' \tau '}|^2$ are expected to give very small contributions,%
\footnote{Unless $|\mu_{e'\mu'}|^2$ is anomalously small.}
 since they are proportional to $s^2_{13}/\Delta^2$. 

There are three main differences between this result and that obtained in ref.~\cite{Akhmedov:2002mf} in the 2-flavour approach. 
First, the main (first) term in (\ref{eqn:peebar}) contains an additional factor $c^4_{13}$. Second, there is a cross-term contribution in (\ref{eqn:peebar}) which is absent from the two-flavour result and which may give rise to a non-negligible correction to the $\bar{\nu}_{eR}$ appearance probability.
Finally, the expression in eq.~(\ref{eqn:peebar}) can be used 
for neutrino energies below $\sim 5 - 8$ MeV, for which the analytical two-flavour result of \cite{Akhmedov:2002mf} is not applicable because of simplifying assumptions made. 

It is interesting to note that the main term in (\ref{eqn:peebar}) is proportional to $|\mu_{e'\mu'}|^2$ which is equal to $|\mu_{12}|^2$, i.e.\ the electron antineutrino appearance probability is, to a very good approximation, proportional to $|\mu_{12}B_\perp (r_0)|^2$.

It will be shown in section~\ref{sec:compare} that the $\bar{\nu}_{eR}$ appearance probability (\ref{eqn:peebar}) is a relatively slowly varying function of neutrino energy for $E\gtrsim 5 - 8$ MeV, relevant for experiments on detection of solar $^8$B neutrinos. Taking for an estimate its value at $E=12$ MeV and electron and neutron number densities at the neutrino production point $N_e \simeq 89$\,cm$^{-3}$ and $N_n \simeq 35$\,cm$^{-3}$ \cite{Serenelli:2009yc}, the electron antineutrino appearance probability can be written as 
\begin{equation}
P (\nu_{eL} \rightarrow \bar{\nu}_{eR}) \simeq 1.1\times 10^{-10} \left(
\frac{\mu_{12}B_\perp (r_0)}{10^{-12}\mu_B \cdot 10\, {\rm kG}}\right)^2\,,
\label{eq:analit1}
\end{equation}
where $\mu_B$ is the electron Bohr magneton. 
The numerical coefficient here is about a factor of 1.4 smaller 
than it is in the two-flavour approach of ref.~\cite{Akhmedov:2002mf}) 
(see eq.~(25) of that paper). This is partly due to 3-flavour effects and to 
using the updated neutrino mixing parameters and solar models and partly 
because of the approximation $\cos\tilde{\theta}(r_0)\ll\sin\tilde{\theta}
(r_0)$ adopted in~\cite{Akhmedov:2002mf} (see  section~\ref{sec:compare} below). 

Note that eq.~(\ref{eq:analit1}) is not suitable for experiments sensitive to pp, pep or $^7$Be solar neutrinos, for which the electron antineutrino appearance probability is strongly suppressed (at $E\sim 1$ MeV, it is approximately three orders of magnitude smaller than that given by eq.~(\ref{eq:analit1})) and also exhibits a stronger energy dependence. We will discuss this issue in more detail in section~\ref{sec:compare}. 

\subsection{\label{sec:num}Numerical calculations}

Instead of developing an approximate analytical solution, one can solve the complete set of the six coupled evolution equations (\ref{eq:evol3}) numerically, tracing the evolution of the system from the neutrino production point in the Sun to the Earth. 
We have developed a numerical code to calculate the electron antineutrino appearance probability at the surface of the Earth for $^8$B neutrinos, which give the main contribution to the solar $\bar{\nu}_{eR}$ flux for energies above the 
threshold of inverse beta decay on protons, i.e. for $E > 1.8$ MeV.%
\footnote{
We are mostly interested in inverse beta decay because the best currently available limits have been obtained from the experiments that used this detection channel.}
The calculations average over the production region of $^8$B neutrinos in the Sun and take into account the electron and neutron number density profiles in the Sun. 
We have performed the calculations for two standard solar models (SSM) which differ on the elemental abundances in the Sun and hence have different metallicities. 
These are the high-metallicity GS98 model and the low-metallicity AGSS09 model, as discussed in \cite{Serenelli:2009yc}. 
As the profile of the magnetic field inside the Sun is essentially unknown, one has to resort to model field profiles, which are actually rather arbitrary. Fortunately, this arbitrariness is to some extent alleviated by the fact that one expects the $\bar{\nu}_e$ appearance probability to be mostly sensitive to the magnetic field strength at the neutrino production point rather than to the complete profile. We test this numerically by using three different magnetic field profiles that have the same strength at $r=0.05R_\odot$ (see section \ref{sec:profile} below).
If not otherwise specified, in our calculations we will be 
using the linearly decreasing magnetic field profile inside the Sun  
\begin{equation}
B_\perp(r)=B_0(r) \equiv 52600 (1 - r/R_\odot)\,{\rm kG}\,, 
\label{eq:b0}
\end{equation}
which takes the value 
$B_\perp \simeq 5 \times 10^{7}$\,G 
at $r_0$ = 0.05R$_\odot$ and vanishes at the surface of the Sun. Possible twist of the solar magnetic field will be neglected. 
The magnetic field profile \eqref{eq:b0} coincides with the one used in \cite{Akhmedov:2002mf}.

\section{Results}
\label{sec:results}

\subsection{\label{sec:compare}Comparison between analytical and numerical results}

We have shown that the main contribution to the electron antineutrino appearance probability is proportional to $|\mu_{e'\mu'}|^2=|\mu_{12}|^2$. In this subsection we set $\mu_{13} =\mu_{23}=0$ (which also implies $\mu_{\mu'\tau'}=\mu_{e'\tau'}=0$) 
and compare our analytical expressions with the results obtained by numerical solution of the system of the evolution equations (\ref{eq:evol3}). 
With only $\mu_{e'\mu'}$ different from zero, our analytical expression (\ref{eqn:peebar}) becomes  
\begin{align}
   P(\nu_{eL} \rightarrow \bar{\nu}_{eR})  = \frac{1}{2} c^4_{13}\sin^ 2 
2\theta_{12}B^2_\perp (r_0)|\mu_{e'\mu'}|^2 \Bigg(\frac{\cos^2
\tilde{\theta}(r_0)}{g'_1(r_0)} + \frac{\sin^2\tilde{\theta}(r_0)}
{g'_2(r_0)}\Bigg) ^ 2.
        \label{eqn:peebar_full}
    \end{align}
We will also consider the simplified analytical expression  
    \begin{align}
        P (\nu_{eL} \rightarrow \bar{\nu}_{eR})_{\rm simpl.}  = 
\frac{1}{2} c^4_{13}\sin^2 2\theta_{12}B^2_\perp (r_0)|\mu_{e'\mu'}|^2 
\Bigg( \frac{\sin^2\tilde{\theta}(r_0)}{g'_2(r_0)}\Bigg) ^ 2, 
        \label{eqn:peebar_simple}
    \end{align}
obtained from (\ref{eqn:peebar_full}) by neglecting the first term in the brackets compared to the second one, the approximation similar to the one adopted in \cite{Akhmedov:2002mf}.  
This approximation is expected to be valid for relatively high neutrino energies, for which $\cos^2\tilde{\theta}(r_0)\ll \sin^2\tilde{\theta}(r_0)$ (note that $|g_1'(r_0)|$ and $|g_2'(r_0)|$ differ by less than a factor of two for all considered energies).
\begin{figure}[t!]
    \centering
    \includegraphics[width = 0.48\textwidth]{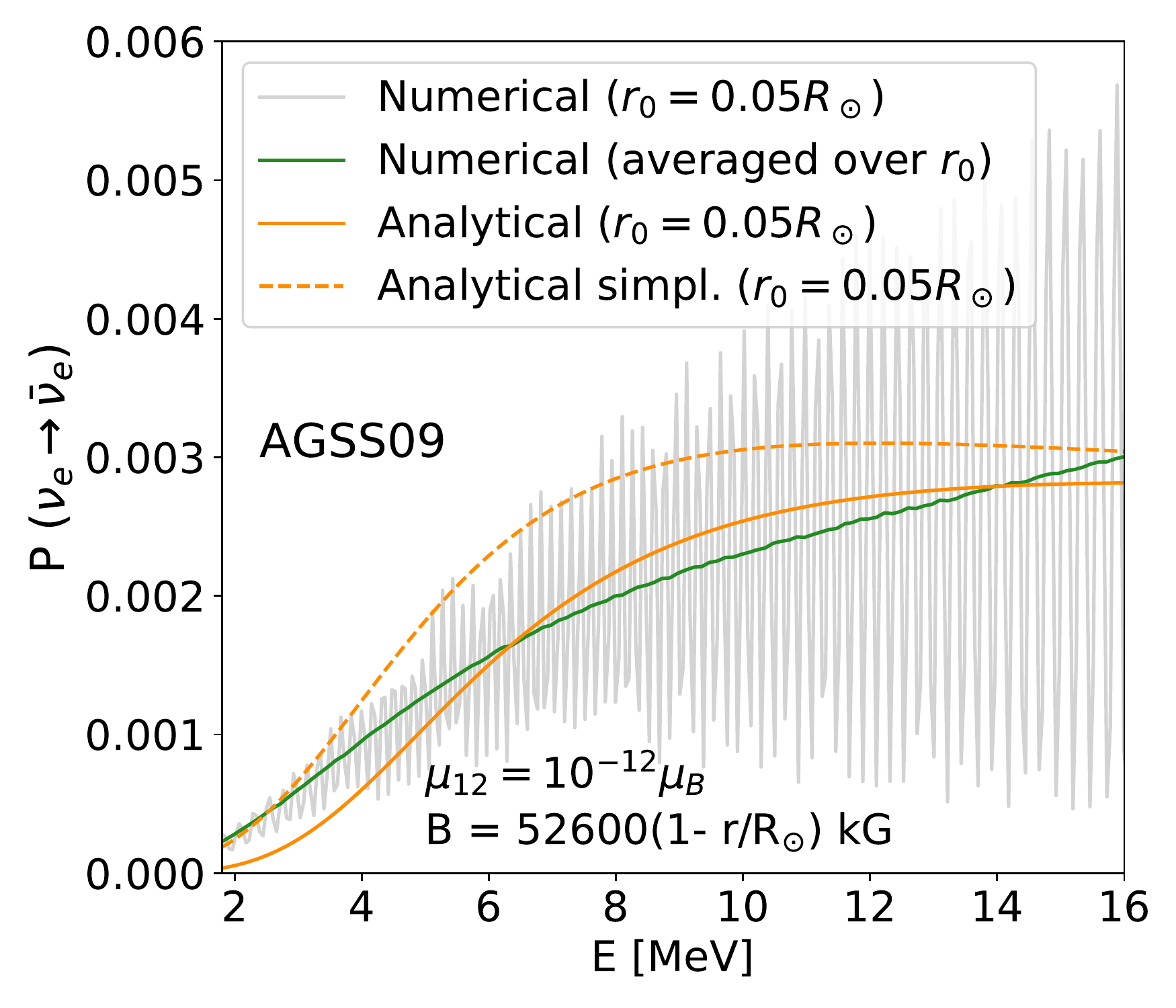}
    \includegraphics[width = 0.48\textwidth]{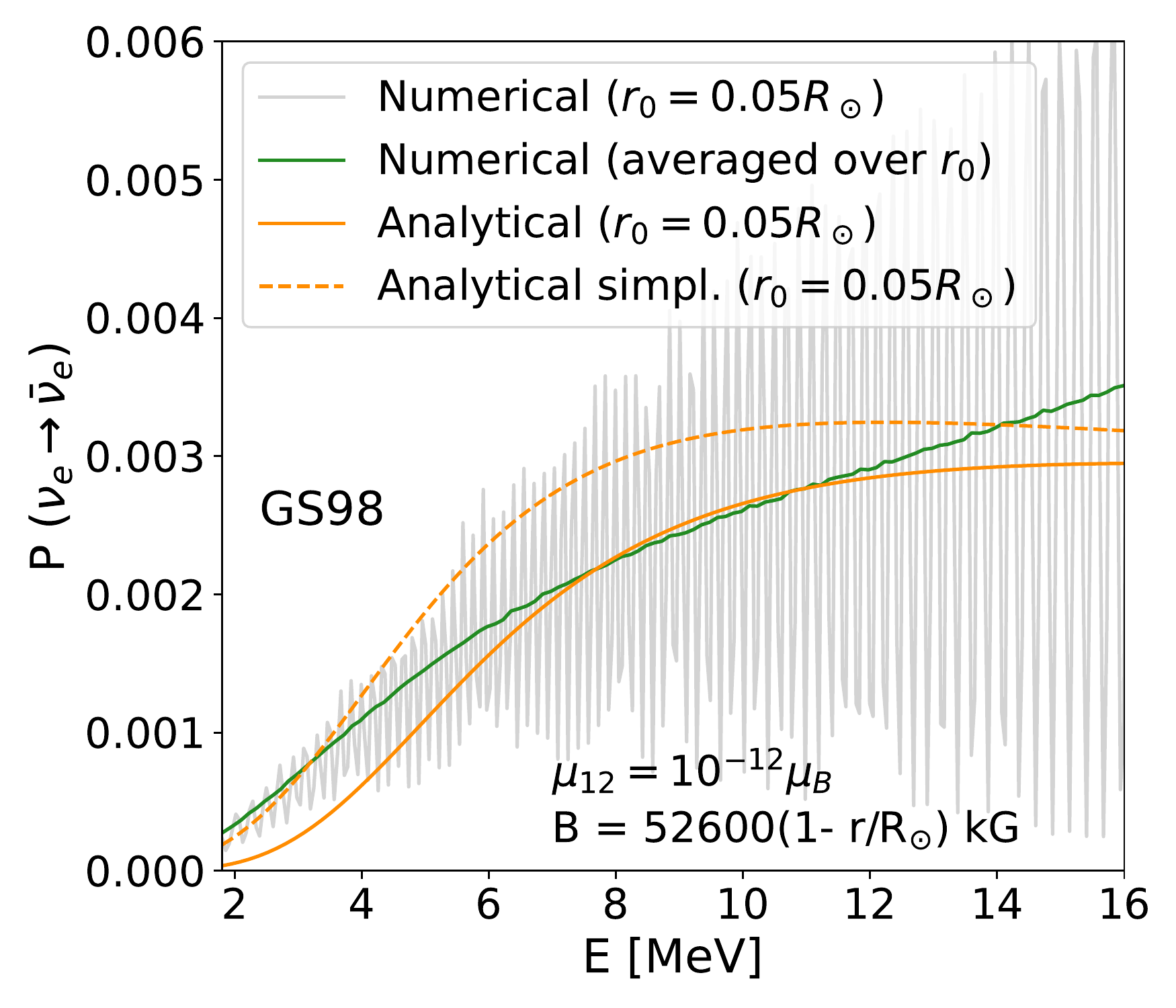}
    \caption{Comparison of different calculations for $\bar{\nu}_{eR}$ appearance probability for $^8$B neutrinos. Values of the oscillation parameters ($\sin^2\theta_{12}$, $\Delta m^2_{21}$) = (0.32, 7.5$\times 10^{-5}$ eV$^2$) and $\sin^2\theta_{13} = 0.022$ \cite{deSalas:2020pgw} were chosen. Left and right panels correspond to AGSS09 and GS98 SSM, respectively \cite{Serenelli:2009yc}. Grey curves: numerical calculations assuming that all neutrinos are produced at $r_0 =0.05 R_\odot$. 
Green curves: numerical calculation with averaging over the neutrino production region. Orange curves: results based on the full analytical expression \eqref{eqn:peebar_full} (solid) and  on the simplified analytical expression \eqref{eqn:peebar_simple}(dashed).
}
    \label{fig:Peebar2020}
\end{figure}

In Figure \ref{fig:Peebar2020} we compare our analytical results with those found by numerical solution of the neutrino evolution equations (\ref{eq:evol3}). The grey wiggly curves show the numerical results obtained assuming that all neutrinos are produced at the distance $r_0=0.05 R_\odot$ from the centre of the Sun; the wiggly behaviour gets washed out if one averages over the neutrino production region, as shown by the green curves.
The solid and dashed orange curves correspond to the analytical expressions \eqref{eqn:peebar_full} and \eqref{eqn:peebar_simple}, respectively, assuming that all neutrinos are produced in the Sun at $r_0=0.05R_\odot$. 
The left and right panels show the results for AGSS09 and GS98 solar models, respectively. The figure demonstrates a good general agreement between our numerical and analytical results, especially for neutrino energies $E\gtrsim 5$ MeV. The discrepancy between the numerical and analytical results becomes larger for smaller $E$, where the $\bar{\nu}_{eR}$ appearance probability is relatively small.   

\subsection{Neutrino evolution inside the Sun}
In order to gain a better insight into the process of anitneutrino appearance, we consider the evolution of the neutrino system inside the Sun as a function of coordinate. For simplicity, we do so in the effective 2-neutrino approach, which corresponds to setting $s_{13}\to 0$. We have checked that the obtained results give a good approximation to those in the full 3-flavour case, the reason being that the corresponding corrections are of the order of $s_{13}^2$.

\begin{figure}
\includegraphics[width = 0.98\textwidth]{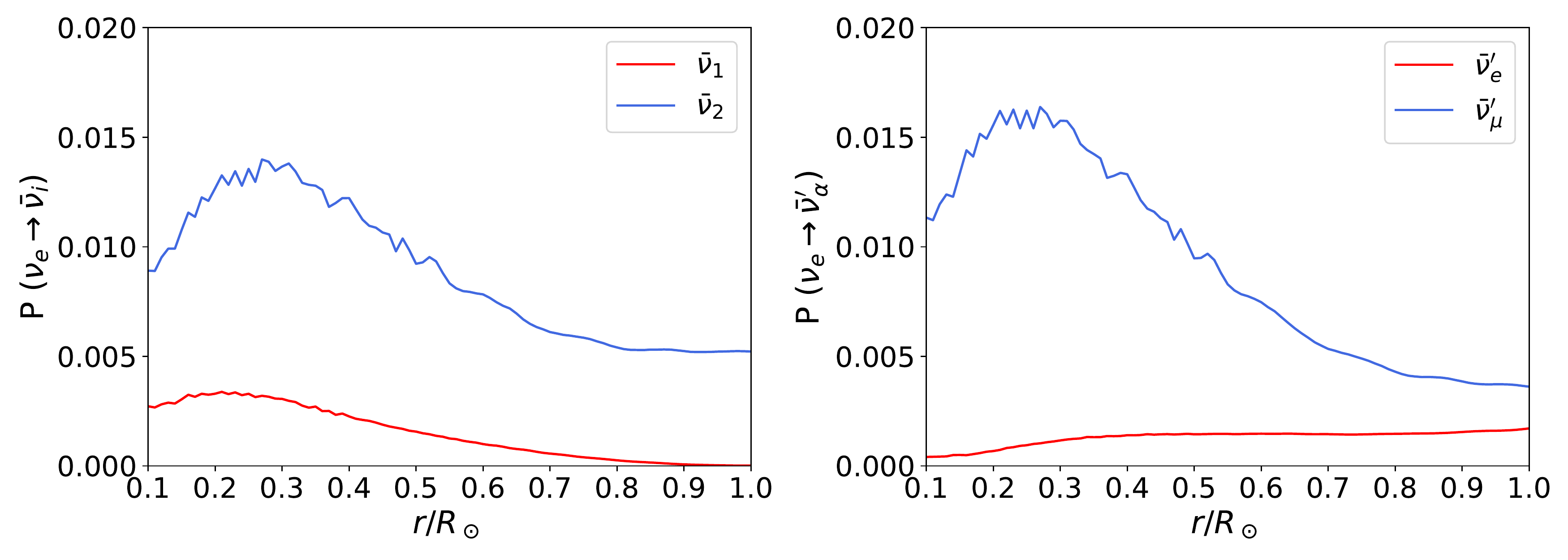}
    \caption{
Two-flavour antineutrino appearance probabilities for the 
mass eigenstates (left panel) and for the states in the primed basis (right panel) as functions of the distance to the centre of the Sun. Neutrino energy $E=10$ MeV, transition magnetic moments $\mu_{12}= 10^{-12} \mu_B$, $\mu_{13} = \mu_{23}=0$ and the $\nu_{e}$ production coordinate $r_0= 0.05R_\odot$ are chosen.}
    \label{fig:inside-bases}
\end{figure}

In Figure \ref{fig:inside-bases} we show the evolution of the antineutrino appearance probabilities, obtained by numerical solution of the evolution equations, for mass-eigenstate (left panel) and primed (right panel) neutrino states. 
From the right panel, one can see that the approximation $\bar{\nu}_e'\sim 0$ is reasonably good inside the Sun, though it is less accurate at its surface. 

\begin{figure}
\includegraphics[width = 0.98\textwidth]{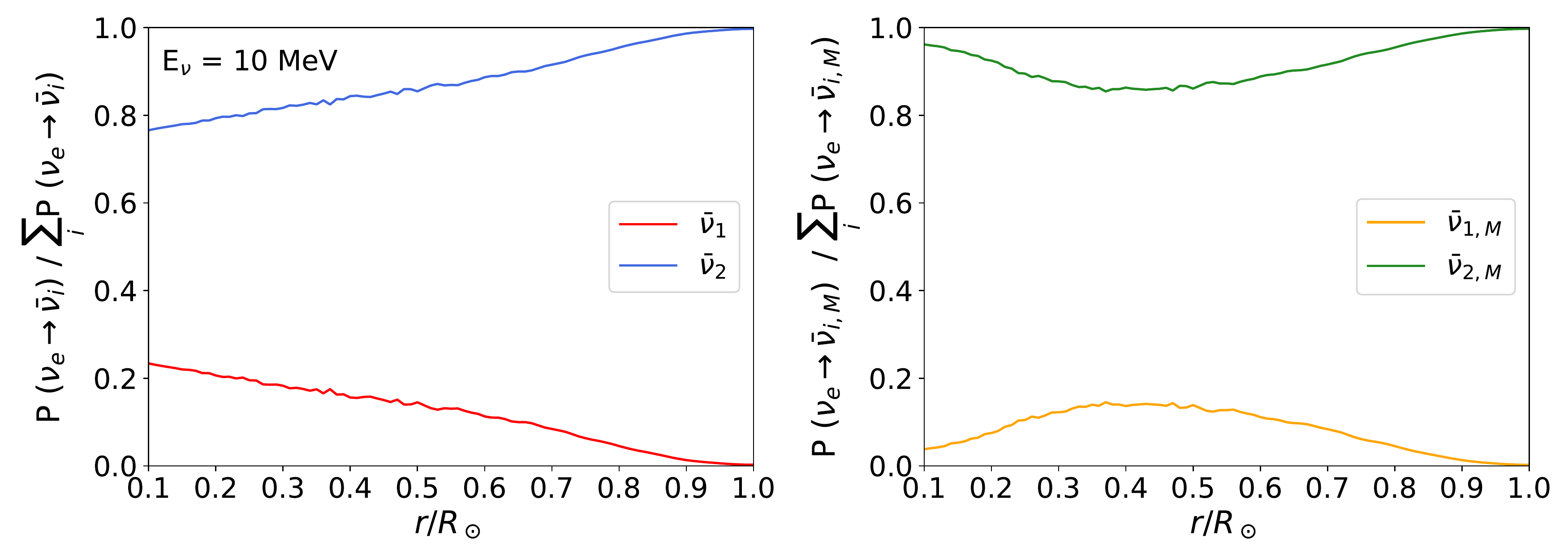}
    \caption{Two-flavour evolution of antineutrino appearance probabilities inside the Sun for mass eigenstates (left panel) and for the matter eigenstates (right panel), normalised to the unit total antineutrino appearance probability. Assumptions regarding neutrino energy, magnetic moments and the $\nu_e$ production coordinate are the same as in Figure 
    \ref{fig:inside-bases}.}
    \label{fig:inside-norm}
\end{figure}

In the two-flavour approximation, SFP converts $\nu_{e}$ produced in the Sun into $\bar{\nu}_\mu'$. Close to the neutrino production point the composition of $\bar{\nu}_\mu'$ is approximately given by \mbox{ $\bar{\nu}_\mu'\simeq - \sin \tilde{\bar{\theta}}(r_0) \bar{\nu}_{1M} + \cos \tilde{\bar{\theta}}(r_0) \bar{\nu}_{2M}$,} where $\bar{\nu}_{1,2 M}$ are antineutrino matter eigenstates (i.e.\ the states that diagonalize the antineutrino Hamiltonian in matter), and the mixing angle $\tilde{\bar{\theta}}(r)$ is given by   
\begin{equation}
\tan 2\tilde{\bar{\theta}}(r) = \frac{\sin 2 \theta_{12}}{\cos 2\theta_{12} 
+ c^2_{13}V_e(r)/2\delta}\, .
\end{equation}
Close to the neutrino production point the electron number density is rather large, and for neutrino energies $E \gtrsim 2 \text{ MeV}$ one has $\tilde{\bar{\theta}}(r_0)\ll 1$, so that $\bar{\nu}'_\mu \simeq \bar{\nu}_{2M}$. As there is no level crossing for antineutrinos and, in addition, their evolution is adiabatic in the Sun, $\bar{\nu}_{2M}$ propagate through the Sun without noticeable transformations to $\bar{\nu}_{1M}$.
Because matter density essentially vanishes at $r=R_\odot$, matter eigenstates become mass eigenstates there, and therefore antineutrinos emerge at the surface of the Sun as $\bar{\nu}_{2}$. This can be seen in Figure \ref{fig:inside-norm}, where we show the appearance probabilities for $\bar{\nu}_1$ and $ \bar{\nu}_2$ (left panel) and for the matter eigenstates $\bar{\nu}_{1M}$ and $ \bar{\nu}_{2M}$ (right panel), normalised to the unit sum.  
For $r=0.1R_{\odot}$, which is relatively close to the neutrino production point, most of the antineutrinos are $\bar{\nu}_{2M}$, which is a nontrivial combination of $\bar{\nu}_{1}$ and $\bar{\nu}_2$. 
At the surface of the Sun, the antineutrinos emerge as $\bar{\nu}_{2M}$ as well, which coincides there with $\bar{\nu}_2$. This is in accord with left panel of Figure \ref{fig:inside-bases}, which shows that at $r=R_\odot$ we mainly find $\bar{\nu}_2$. 
As $\bar{\nu}_2$ is a linear combination of $\bar{\nu}_e'$ and $\bar{\nu}_\mu'$ with weights $\sin^2\theta_{12}\simeq 1/3$ and $\cos^2\theta_{12}\simeq 2/3$ respectively, at the surface of the Sun the appearance probability of $\bar{\nu}_\mu'$ is about twice that of $\bar{\nu}_e'$ (right panel of Figure \ref{fig:inside-bases}). 
It should be noted that, unlike for the normalised probabilities shown in Figure \ref{fig:inside-norm}, the sum of the antineutrino appearance probabilities presented in Figure \ref{fig:inside-bases} is not conserved; this is due to the fact that some of antineutrinos can precess back to neutrinos in the course of their evolution inside the Sun. 
 
\subsection{The roles of various transition magnetic moments and of the magnetic field profile}
\label{sec:profile}

\begin{figure}[t]
\centering
\includegraphics[width = 0.6\textwidth]
{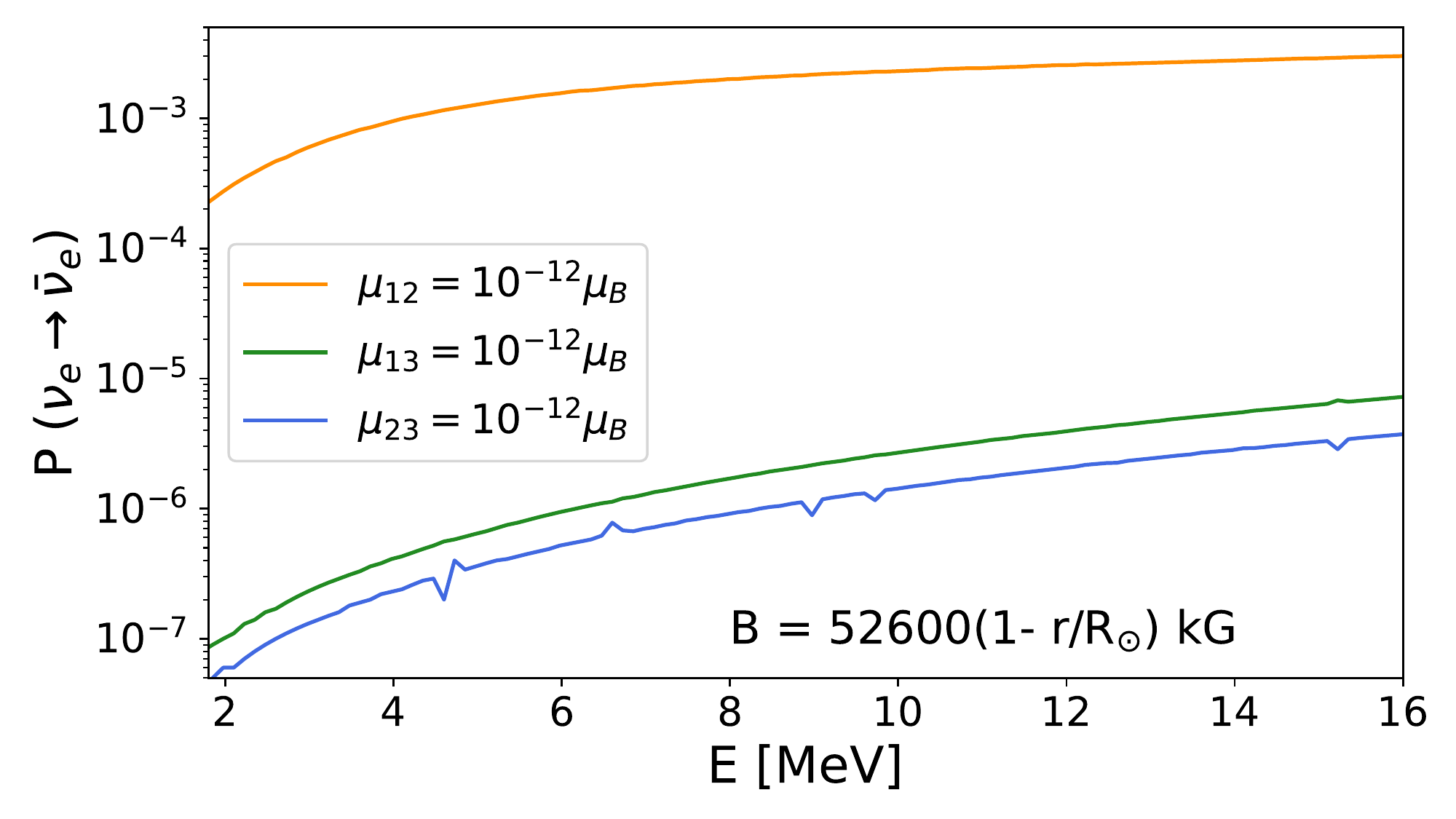}
    \caption{Electron antineutrino appearance probability on the Earth as a function of neutrino energy for nonzero $\mu_{12}$, $\mu_{13}$ and $\mu_{23}$, for the AGSS09 solar model. Evolution equations were solved numerically.}
    \label{fig:comparison}
\end{figure}

Up to this point, in our numerical analysis we were assuming only one transition magnetic moment, $\mu_{e'\mu'}=\mu_{12}$, to be nonzero. This was motivated by our analytical results, which showed that the contributions $\mu_{e'\tau'}$ and $\mu_{\mu'\tau'}$, which are linear combinations $\mu_{13}$ and $\mu_{23}$, are strongly suppressed. 
To illustrate this point, in Figure \ref{fig:comparison} we present the $\bar{\nu}_e$ appearance probability $P(\nu_e\to \bar{\nu}_e)$ at the Earth when one nonzero magnetic moment at a time is allowed. It clearly demonstrates that, unless $\mu_{13}$ or $\mu_{23}$ are more than three orders of magnitude larger than $\mu_{12}$, the latter completely dominates the $\nu_e\to \bar{\nu}_e$ conversion.  

In the above, all the numerical results were obtained for the simple linear model magnetic field of the Sun $B_0(r)$ (\ref{eq:b0}). We shall now study the sensitivity of $\nu_e\to \bar{\nu}_e$ conversion to the solar magnetic field profile.  
To this end, we compare the results obtained for the linear profile we used above with those for two different parabolic profiles, $B_1(r)$ and $B_2(r)$. All the profiles are chosen to have the same strength $5\times 10^4$\,kG at $r=0.05R_\odot$ and to vanish at the surface of the Sun. 
The profile 
    \begin{equation}
    B_1(r) = 50000 + 2632 \frac{r}{R_\odot} - 52632 \left(\frac{r}
    {R_\odot}\right)^2 \, {\rm kG} \, 
    \label{eqn:b1}
    \end{equation}
is almost flat over the production region; the profile  
    \begin{equation}
    B_2(r) = 55000 - 102368 \frac{r}{R_\odot} + 47368 \left(\frac{r}
    {R_\odot}\right)^2 \, {\rm kG} \,.
    \label{eqn:b2}
    \end{equation} 
corresponds to the magnetic field that is smaller than the linear one for $r > 0.05R_\odot$. 

\begin{figure}
    \centering
    \includegraphics[width = 0.99\textwidth]
{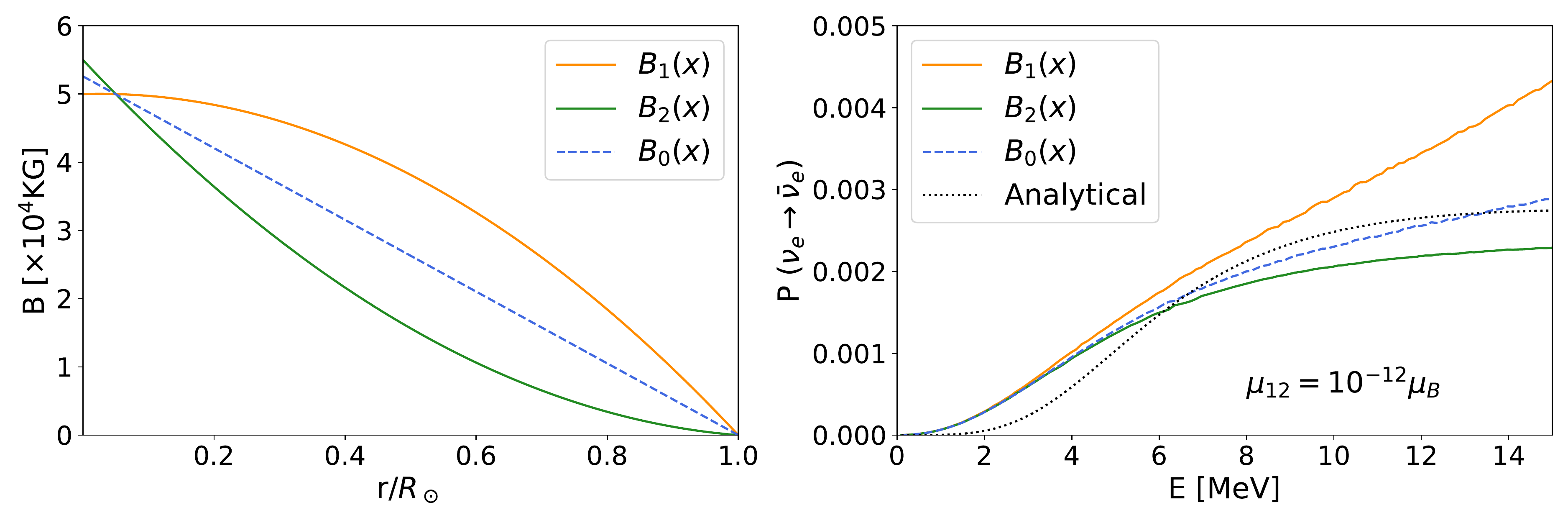}
    \caption{Left panel: model magnetic field profiles inside the Sun, 
as defined in eqs.~(\ref{eq:b0}), (\ref{eqn:b1}) and (\ref{eqn:b2}). 
Right panel: the corresponding electron antineutrino appearance probabilities on the Earth. Results of the analytical expression shown by black dotted curve; for the rest of the curves evolution equations were solved numerically. }
    \label{fig:magn_profile}
\end{figure}

In Figure \ref{fig:magn_profile} (left panel) we plot the magnetic field profiles we use.  In the right panel the corresponding $\bar{\nu}_e$ appearance probabilities are shown. For neutrino energies $E\lesssim 7$\,MeV all the employed magnetic field profiles lead to $\bar{\nu}_e$ appearance probabilities that are quite close to each other. The sensitivity to the magnetic field profile increases with neutrino energy. 
The reason for this is twofold. First, neutrinos are produced in the Sun not at the same distance from its centre (such as e.g. $0.05R_\odot$ which we considered as a reference value for our estimates and where all our model magnetic field profiles coincide), but their production actually takes place over the extended region; the $\nu_e\to \bar{\nu}_e$ production probability is therefore sensitive to the magnetic profile in that region. 
Second, the $\bar{\nu}_e$ appearance probability depends on the ``mixing'' of the left-handed and right handed neutrinos at their production point $r_0$, which is proportional to $\mu_{12}B_\perp(r_0)/(\Delta m_{21}^2/2E)$, which increases with neutrino energy. 
From Figure \ref{fig:magn_profile} it follows that for $E\sim 8$\,MeV (which is a typical energy of $^8$B neutrinos) one can expect the sensitivity of the $\bar{\nu}_e$ appearance probability to the choice of the magnetic field profile to be of the order of 10 $-$ 15\%. 

\subsection{Average electron antineutrino appearance probability and expected flux}
A number of experimental collaborations have reported upper limits on $\bar{\nu}_e$ flux from astrophysical sources. These are obtained for certain energy ranges as
\begin{equation}
    \Phi_{\rm C.L.} = \frac{N_{\rm C.L.}}{\epsilon \cdot \langle\sigma\rangle
\cdot T\cdot N_p} \, ,
\end{equation}
where $N_{\rm C.L.}$ is upper limit on the number of events at a given confidence level, $\epsilon$ is the average detection efficiency in the energy range considered, $\langle\sigma\rangle$ is the averaged cross-section in the same energy range, $T$ is the exposure time and $N_p$ is the number of target particles.

In order to make it easier to use our results for analyses of the existing and future experimental data, we compute, for each energy bin $E \in [E_i-\Delta E/2, E_i+\Delta E/2]$, the averaged $\bar{\nu}_e$ appearance probability 
\begin{equation}
    \langle P_{i} \rangle = \frac{ \bigintss_{E_i - \Delta E/2}^{E_i 
+\Delta E/2} \phi(E) \sigma(E) P(E) dE}{ 
\bigintss_{E_i - \Delta E/2}^{E_i +\Delta E/2} \phi(E) \sigma(E) dE} \, ,
\label{eq:Pi}
\end{equation}
and the expected $\bar{\nu}_e$ flux  
\begin{equation}
    \langle\Phi_{i}\rangle = \frac{ \bigintss_{E_i - \Delta E/2}^{E_i +
\Delta E/2} \phi(E) \sigma(E) P(E) dE}{ \bigintss_{E_i - 
\Delta E/2}^{E_i +\Delta E/2} \sigma(E) dE} \,.
\label{eq:Phii}
\end{equation}
For simplicity, we have assumed that the detection efficiency $\epsilon$ is nearly energy independent within each bin (though it may vary from bin to bin); it then cancels out in the ratios (\ref{eq:Pi}) and (\ref{eq:Phii}).
We also assumed perfect detector energy resolution; we have checked that for the KamLAND experiment, taking into account the realistic energy resolution of 6.4\%/$\sqrt{E {\rm (MeV)}}$, changes our results by less than 0.5\%. This is related to the fact that the $\bar{\nu}_e$ appearance probability is a rather smooth function of neutrino energy (see the right panel of Figure~\ref{fig:magn_profile}). 

We restrict ourselves to energies above 1.8 MeV, where only $^8$B neutrinos give a significant contribution to the solar neutrino signal, and we consider the inverse beta decay as the $\bar{\nu}_e$ detection process.
We compute the $\bar{\nu}_e$ appearance probability and the expected flux numerically, using both the numerical and analytical expressions for the probabilities $P(\nu_e\to\bar{\nu}_e)$.
In Tables \ref{tab:estimatesAGSS09} and \ref{tab:estimatesGS98} we present these probabilities and the expected $\bar{\nu}_e$ fluxes for the fixed values $\mu_{12} = 10^{-12} \mu_B$ and $B_\perp(r_0) = 1$ kG. As the $\bar{\nu}_e$ appearance probability and the $\bar{\nu}_e$ flux are proportional to $(\mu_{12}B_\perp(r_0))^2$, the values of $\langle P_i\rangle$ and $\langle \Phi_i\rangle$ for different $\mu_{12}B_\perp(r_0)$ can be found by simple rescaling.

\begin{table}[t]
    \centering
    \renewcommand{\arraystretch}{1.4}
	\begin{tabular}{c|c|c|c|c|} \cline{2-5}
 & \multicolumn{2}{c|}{Numerical AGSS09} & \multicolumn{2}{c|}
{Analytical AGSS09} 
\\ \hline
\multicolumn{1}{|c|}{E [MeV]} &  $\langle P_{i} \rangle$ & $\langle\Phi_{i}
\rangle [\text{cm}^{-2}  s^{-1}  \text{MeV}^{-1}]$ & $\langle P_{i} 
\rangle$ & $\langle\Phi_{i}\rangle [\text{cm}^{-2}  s^{-1}  \text{MeV}^{-1}]$  
 \\ \hline \hline 
\multicolumn{1}{|c|}{1.8	- 2.8}	&	1.73$\times 10^{-13}$	&
4.62$\times 10^{-8}$	&	5.08$\times 10^{-14}$	&	1.36$\times 
10^{-8}$	\\
\multicolumn{1}{|c|}{2.8	- 3.8}	&	2.95$\times 10^{-13}$	&
1.18$\times 10^{-7}$	&	1.45$\times 10^{-13}$	&	5.82$\times 
10^{-8}$	\\
\multicolumn{1}{|c|}{3.8	- 4.8}	&	4.28$\times 10^{-13}$	&
2.24$\times 10^{-7}$	&	2.97$\times 10^{-13}$	&	1.55$
\times 10^{-7}$	\\
\multicolumn{1}{|c|}{4.8	- 5.8}	&	5.52$\times 10^{-13}$	&
3.34$\times 10^{-7}$	&	4.73$\times 10^{-13}$	&	2.86$
\times 10^{-7}$	\\
\multicolumn{1}{|c|}{5.8	- 6.8}	&	6.57$\times 10^{-13}$	&
4.19$\times 10^{-7}$	&	6.38$\times 10^{-13}$	&	4.07$
\times 10^{-7}$	\\
\multicolumn{1}{|c|}{6.8	- 7.8}	&	7.45$\times 10^{-13}$	&
4.62$\times 10^{-7}$	&	7.74$\times 10^{-13}$	&	
4.80$\times 10^{-7}$	\\
\multicolumn{1}{|c|}{7.8	- 8.8}	&	8.19$\times 10^{-13}$	&
4.55$\times 10^{-7}$	&	8.78$\times 10^{-13}$	&	
4.88$\times 10^{-7}$	\\
\multicolumn{1}{|c|}{8.8	- 9.8}	&	8.84$\times 10^{-13}$	&
4.03$\times 10^{-7}$	&	9.53$\times 10^{-13}$	&	
4.35$\times 10^{-7}$	\\
\multicolumn{1}{|c|}{9.8  - 10.8}	&	9.38$\times 10^{-13}$	&
3.15$\times 10^{-7}$	&	1.01$\times 10^{-12}$	&	
3.38$\times 10^{-7}$	\\
\multicolumn{1}{|c|}{10.8 - 11.8}	&	9.87$\times 10^{-13}$	&
2.11$\times 10^{-7}$	&	1.04$\times 10^{-12}$	&	
2.23$\times 10^{-7}$	\\
\multicolumn{1}{|c|}{11.8 - 12.8}	&	1.04$\times 10^{-12}$	&
1.12$\times 10^{-7}$	&	1.07$\times 10^{-12}$	&	
1.16$\times 10^{-7}$	\\
\multicolumn{1}{|c|}{12.8 - 13.8}	&	1.08$\times 10^{-12}$	&
3.87$\times 10^{-8}$	&	1.08$\times 10^{-12}$	&	
3.89$\times 10^{-8}$	\\
\multicolumn{1}{|c|}{13.8 - 14.8}	&	1.12$\times 10^{-12}$	&
5.77$\times 10^{-9}$	&	1.09$\times 10^{-12}$	&	
5.63$\times 10^{-9}$	\\
\multicolumn{1}{|c|}{14.8 - 15.8}	&	1.16$\times 10^{-12}$	&
2.83$\times 10^{-10}$	&	1.10$\times 10^{-12}$	&	
2.68$\times 10^{-10}$	\\ \hline
\end{tabular}
    \caption{\label{tab:estimatesAGSS09} Averaged $\bar{\nu}_e$ appearance probabilities and expected fluxes of $\bar{\nu}_e$ from the Sun for low-metallicity AGSS09 SSM. Detection through inverse beta decay is assumed; magnetic field profile (\ref{eq:b0}) and $\mu_{12}B_\perp(r_0) =10^{-12}\mu_B 
\cdot {\rm kG}$ were chosen. 
For rescaling to different values of $\mu_{12}B_\perp(r_0)$ see text.
}
\end{table}

\begin{table}[t]
    \centering
    \renewcommand{\arraystretch}{1.4}
	\begin{tabular}{c|c|c|c|c|} \cline{2-5}
 & \multicolumn{2}{c|}{Numerical GS98} & \multicolumn{2}{c|}{Analytical GS98} 
\\ \hline
\multicolumn{1}{|c|}{E [MeV]} &  $\langle P_{i} \rangle$ & $\langle\Phi_{i}
\rangle [\text{cm}^{-2}  s^{-1}  \text{MeV}^{-1}]$ & $\langle P_{i} \rangle$ 
& $\langle\Phi_{i}\rangle [\text{cm}^{-2}  s^{-1}  \text{MeV}^{-1}]$    
\\ \hline \hline 
\multicolumn{1}{|c|}{1.8	-2.8}	& 2.03$\times 10^{-13}$	 & 6.60$\times 10^{-8}$	 &  5.43$\times 10 ^{-14}$	 & 1.76$\times 10^{-8}$	\\
\multicolumn{1}{|c|}{2.8	- 3.8}	& 3.43$\times 10^{-13}$	 & 1.67$\times 10^{-7}$	 &  1.55$\times 10^{-13}$	 & 7.56$\times 10^{-8}$	\\
\multicolumn{1}{|c|}{3.8	- 4.8}	& 4.92$\times 10^{-13}$	 & 3.12$\times 10^{-7}$	 &  3.19$\times 10^{-13}$	 & 2.02$\times 10^{-7}$	\\
\multicolumn{1}{|c|}{4.8	- 5.8}	& 6.27$\times 10^{-13}$	 & 4.60$\times 10^{-7}$	 &  5.06$\times 10^{-13}$	 & 3.72$\times 10^{-7}$	\\
\multicolumn{1}{|c|}{5.8	- 6.8}	& 7.41$\times 10^{-13}$	 & 5.73$\times 10^{-7}$	 &  6.82$\times 10^{-13}$	 & 5.28$\times 10^{-7}$	\\
\multicolumn{1}{|c|}{6.8	- 7.8}	& 8.39$\times 10^{-13}$	 & 6.31$\times 10^{-7}$	 &  8.26$\times 10^{-13}$	 & 6.21$\times 10^{-7}$	\\
\multicolumn{1}{|c|}{7.8 - 8.8}	& 9.23$\times 10^{-13}$	 & 6.22$\times 10^{-7}$	 &  9.35$\times 10^{-13}$	 & 6.30$\times 10^{-7}$	\\
\multicolumn{1}{|c|}{8.8	- 9.8}	& 9.96$\times 10^{-13}$	 & 5.51$\times 10^{-7}$	 &  1.0$\times 10^{-12}$	 & 5.60$\times 10^{-7}$	\\
\multicolumn{1}{|c|}{9.8  - 10.8}	 & 1.06$\times 10^{-12}$  & 4.33$\times 10^{-7}$  &  1.07$\times 10^{-12}$	 & 4.35$\times 10^{-7}$	\\
\multicolumn{1}{|c|}{10.8 - 11.8}	& 1.12$\times 10^{-12}$	 & 2.92$\times 10^{-7}$	 &  1.10$\times 10^{-12}$	 & 2.87$\times 10^{-7}$	\\
\multicolumn{1}{|c|}{11.8 - 12.8}	& 1.18$\times 10^{-12}$	 & 1.55$\times 10^{-7}$	 &  1.13$\times 10^{-12}$	 & 1.48$\times 10^{-7}$	\\
\multicolumn{1}{|c|}{12.8 - 13.8}	& 1.24$\times 10^{-12}$	 & 5.38$\times 10^{-8}$	 &  1.14$\times 10^{-12}$	 & 4.99$\times 10^{-8}$	\\
\multicolumn{1}{|c|}{13.8 - 14.8}	& 1.29$\times 10^{-12}$	 & 8.06$\times 10^{-9}$	 &  1.15$\times 10^{-12}$	 & 7.21$\times 10^{-9}$ \\
\multicolumn{1}{|c|}{14.8 - 15.8}	& 1.34$\times 10^{-12}$	 & 3.98$\times 10^{-10}$	 &  1.16$\times 10^{-12}$	 & 3.43$\times 10^{-10}$ \\ \hline
    \end{tabular}
    \caption{\label{tab:estimatesGS98} Same as in Table \ref{tab:estimatesAGSS09} but for high metallicity GS98 SSM. 
}
\end{table}

\begin{figure}
    \centering
    \includegraphics[width = 0.98\textwidth]{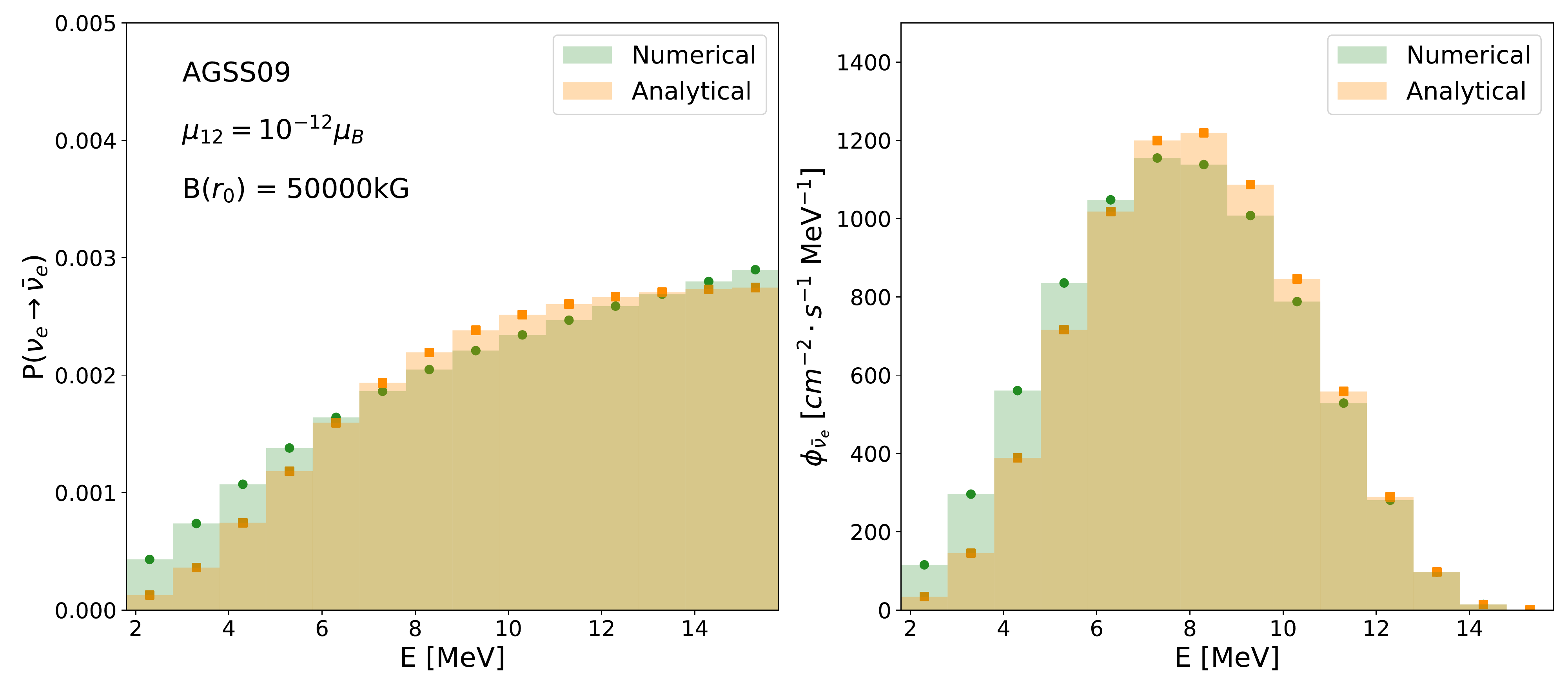}
    \caption{Comparison of the numerical and analytical results               for $\bar{\nu}_e$ appearance for AGSS09 SSM. Left panel: appearance probabilities; right panel: expected $\bar{\nu}_e$ fluxes. Magnetic field of eq.~(\ref{eq:b0}) and $\mu_{12} = 10^{-1 2} \mu_B$ were chosen.}
    \label{fig:pav_flux}
\end{figure}

For better illustration, we also compare in Figure \ref{fig:pav_flux} the $\bar{\nu}_e$ appearance probabilities (left panel) and the predicted $\bar{\nu}_e$ fluxes at the Earth (right panel) obtained numerically and analytically for the case of AGSS09 SSM, magnetic field strength of eq.~(\ref{eq:b0}) and $\mu_{12} = 10^{-12} \mu_B$. 
It can be seen from the figure that for neutrino energies $E\gtrsim 6$ MeV there is a good agreement between our numerical and analytical results; the agreement worsens towards smaller $E$. Thus, while our simple analytical results can be reliably used at relatively high neutrino energies, numerical results should preferably be used for analysing experiments sensitive to low-$E$ part of the solar neutrino spectrum, such as Borexino.     

\subsection{Existing limits from astrophysical $\bar{\nu}_e$ fluxes revisited}

We will revisit here the existing limits on neutrino magnetic moments and solar magnetic fields coming from the upper bounds on astrophysical $\bar{\nu}_e$ fluxes and compare them with our results. 
At present, the most stringent limits come from the KamLAND experiment \cite{KamLAND:2021gvi}, although Borexino and Super-Kamiokande set comparable bounds \cite{Borexino:2019wln, Super-Kamiokande:2020frs,Super-Kamiokande:2021jaq}. In all these experiments the detection channel was inverse beta decay on protons.  
Historically, SNO also put constraints on astrophysical $\bar{\nu}_e$ in the MeV energy range using charge-current interactions with deuterium \cite{SNO:2004eru}, but these limits are not currently competitive.

\begin{figure}
    \centering
    \includegraphics[width = 0.8\textwidth]{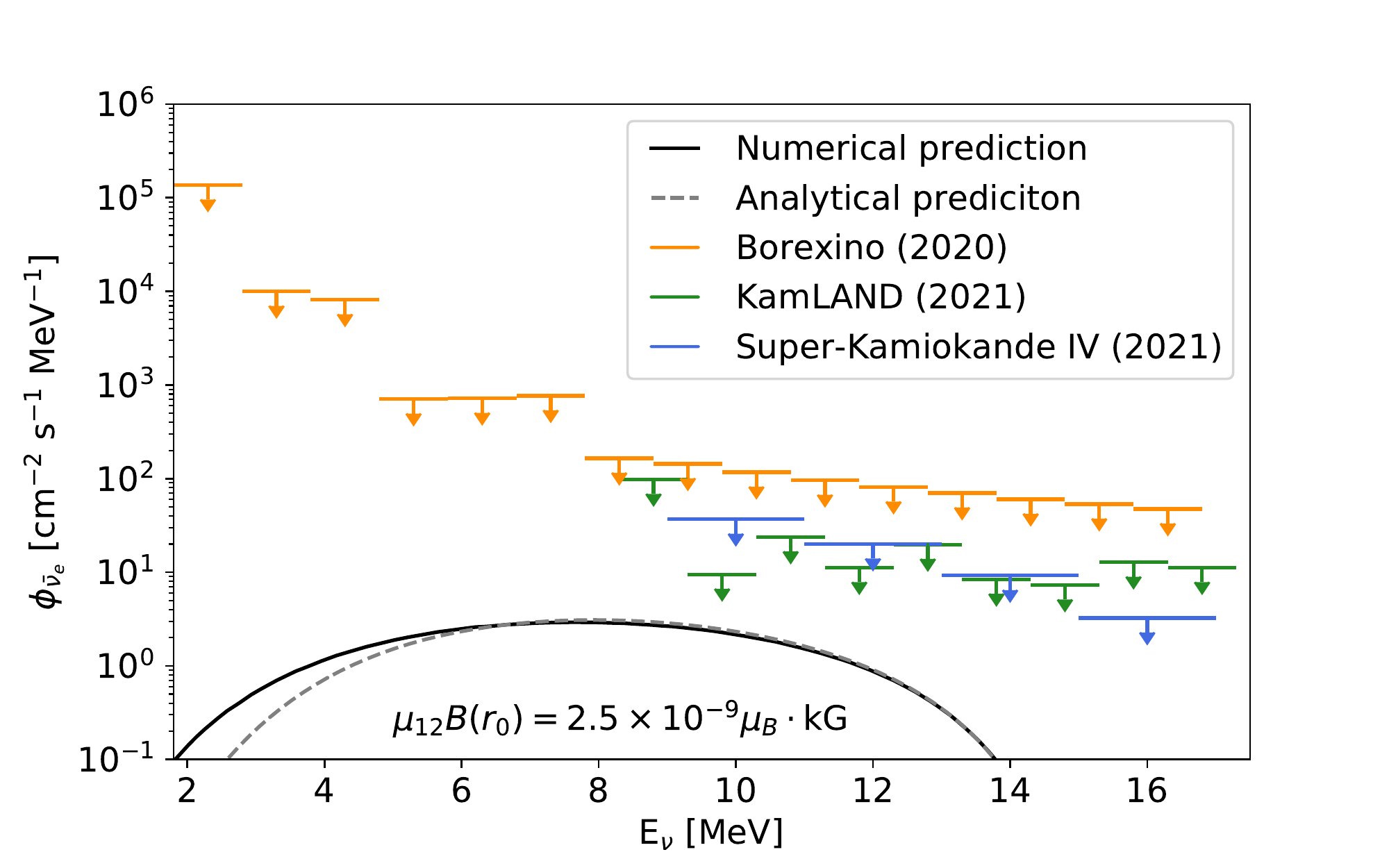}
    \caption{Model-independent limits on $\bar{\nu}_e$ flux of 
    astrophysical origin, as reported by KamLAND \cite{KamLAND:2021gvi}, Borexino \cite{Borexino:2019wln} and Super-Kamiokande \
\cite{Super-Kamiokande:2020frs,Super-Kamiokande:2021jaq}. For comparison, we show the expected solar $\bar{\nu}_e$ flux for $\mu_{12}B_\perp(r_0)= 2.5 \times 10^{-9}\mu_B\,$kG for the AGSS09 SSM, from our analytical and numerical calculations.}
    \label{fig:limits}
\end{figure}

The model-independent limits on the $\bar{\nu}_e$ flux established by the above-mentioned experiments are shown in Figure \ref{fig:limits}, together with our $\bar{\nu}_e$ flux prediction for solar electron antineutrinos for the AGSS09 SSM and for $\mu_{12}B(r_0) = 2.5 \times 10^{-9}\mu_B\,$kG. Notice that the experimental bounds come closest to the predicted flux at neutrino energies $E\sim 10$ MeV. 
Although the experimental bounds are stronger at energies around 20-30 MeV, the flux of solar neutrinos is extremely low for energies above 16 MeV. High-energy experimental bounds may, however, be relevant for constraining the flux of $\bar{\nu}_e$ from supernovae. 

The 90$\%$ C.L. upper limits on the product of the neutrino magnetic moment and the solar magnetic field strength we obtain from the KamLAND upper bound on the astrophysical $\bar{\nu}_e$ flux are, for the two SSM considered, 
\begin{align}
\left(\mu_{12} B_\perp(r_0)\right)_{\rm AGSS09} &< \left( 4.9-5.1\right)
\times 10^{-9} \mu_B\,\text{kG}\,, \nonumber
\\ 
\left(\mu_{12} B_\perp(r_0)\right)_{\rm GS98} &< \left( 4.7 -4.8\right)
\times 10^{-9} \mu_B\,\text{kG}\,.
\label{eqn:KamLANDlimit}
\end{align}
Here the lower numbers correspond to our analytical approximation and the higher ones, to the full numerical calculation. 
A good general agreement between the results of the two approaches can be seen. The obtained results are also consistent with the limits derived in \cite{KamLAND:2021gvi}, $\mu B_\perp(r_0) < 4.9 \times 10^{-9} \mu_B\,{\rm kG}$, where the previous analytical calculation from ref.~\cite{Akhmedov:2002mf} was used.
Similarly, one can derive the 90$\%$ C.L. limits from the Borexino results, 
\begin{align}
\left(\mu_{12} B_\perp(r_0)\right)_{\rm AGSS09} &< \left( 1.8-1.9\right)
\times 10^{-8} \mu_B\,\text{kG}\,, \nonumber
\\ 
\left(\mu_{12} B_\perp(r_0)\right)_{\rm GS98} &< \left( 1.7-1.8\right)\times 
10^{-8} \mu_B\,\text{kG}\,, 
\label{eq:BorLimitOurs}
\end{align}
whereas from the Super-Kamiokande results we find
\begin{align}
\left(\mu_{12} B_\perp(r_0)\right)_{\rm AGSS09} &< \left( 7.1-7.3\right)
\times 10^{-9} \mu_B \, \text{kG} \nonumber
\\ 
\left(\mu_{12} B_\perp(r_0)\right)_{\rm GS98} &< \left( 6.8-6.9\right)
\times 10^{-9} \mu_B \, \text{kG} \, .
\label{eq:SK}
\end{align}
The previously obtained Borexino limit, derived in \cite{Borexino:2019wln} for the high-metallicity GS98 SSM, was $\mu B_\perp(r_0) < 6.9 \times 10^{-9} \mu_B \cdot {\rm kG}$. The factor $\sim 2.6$ discrepancy between this result and our limit (\ref{eq:BorLimitOurs}) is presumably related to the fact that in the Borexino analysis the simplified energy-independent formula (25) from \citep{Akhmedov:2002mf}, derived for $E\sim 5-10$ MeV, was used for neutrinos of smaller energies, i.e.\ outside its range of validity. As a result, Borexino arrived at a more stringent limit.  
 
For the Super-Kamiokande experiment, the limit found in 
\cite{Super-Kamiokande:2020frs}, $\mu B_\perp(r_0) < 1.5 \times 10^{-8} \mu_B \, {\rm kG}$, is approximately a factor 2 weaker than our limit (\ref{eq:SK}). This difference is probably due to the fact that Super-Kamiokande looked for electron antineutrinos in the energy range 9.3 to 17.3 MeV but used in their analysis the same simplified energy-independent $\bar{\nu}_e$ appearance probability that was derived in \citep{Akhmedov:2002mf} for smaller energies.

\section{Other limits on neutrino magnetic moments \label{sec:limits}}

In this section we give an overview of the existing limits on neutrino magnetic moments coming from various experimental searches, paying special attention to the relations between the experimentally accessible quantities and neutrino magnetic moments or their combinations.  
Neutrino magnetic moment contributions to the cross-sections are often parametrised in terms of effective magnetic moments. These quantities depend on Dirac versus Majorana nature of neutrinos, on the flavour of the incoming neutrinos, and may also depend on other experimental details; in particular, flavour transitions on the way between the neutrino source and detector may have to be taken into account. 

We clarify how these effective quantities are related to each other and to the fundamental neutrino magnetic moments in the mass and flavour bases.

\subsection{Limits from electromagnetic contributions to scattering processes}

Photon exchange processes induced by neutrino magnetic moments can affect neutrino scattering processes, such as e.g.\ elastic neutrino-electron scattering (ES) and coherent elastic neutrino-nucleus scattering (CE$\nu$NS). Since the neutrino magnetic dipole moment interactions flip the neutrino chirality while the Standard Model weak interactions conserve it, these contributions add up incoherently. 

Following the formalism in \cite{Grimus:2000tq, Grimus:2002vb}, one can express the effective neutrino magnetic moment as
\begin{equation}
    \mu^2_{\nu_\alpha} = \nu^\dagger_{L}\, (\upmu^\dagger \upmu) \, \nu_{L} + 
\nu^\dagger_{R}\, (\upmu\upmu^ \dagger) \, \nu_{R} \, ,
    \label{eqn:effective_mu}
\end{equation}
where $\nu_{L}$ and $\nu_{R}$ denote the vectors of the amplitudes of the incoming neutrinos with left- and right-handed chiralities, respectively,  and $\upmu$ is the matrix of neutrino magnetic moments. 
For Majorana neutrinos, the transformation between the neutrino amplitudes and magnetic moments in the mass-eigenstate and flavour bases is given by
\begin{equation}
    \nu_{m, L} = U^\dagger \nu_{fl, L} \quad \quad \nu_{m, R} = U^T 
\nu_{fl, R} \quad \quad \upmu_m = U^T\upmu_{fl} U \,,
\label{eq:MajTransf}
\end{equation}
in an obvious notation. For Dirac neutrinos, the mass matrix is in general diagonalised by a bi-unitary transformation with separate rotations for the left-handed and right handed fields, so that the amplitudes and the magnetic moment matrix transform as
\begin{equation}
    \nu_{m,L} = U_{L}^\dagger \nu_{fl, L} \quad \quad \nu_{m, R} = 
U_{R}^\dagger \nu_{fl, R}\,,\quad\quad
\quad \upmu_m = U^ \dagger _{R}  \upmu_{fl} U_{L}
\label{eq:DirTransf}
\end{equation}
{}From eqs.~(\ref{eq:MajTransf}) and (\ref{eq:DirTransf}) it is easy to see that the expression for the effective neutrino magnetic moment in eq.~(\ref{eqn:effective_mu}) is basis-independent, as any observable should be. It is also valid for both Majorana and Dirac neutrino cases, as far as the final-state neutrinos are not detected in the scattering processes.

\subsubsection{Effective neutrino magnetic moments for short-baseline experiments}

For short-baseline scattering experiments, where the distance from the source to the detector is much shorter than the oscillation lengths $L_{ij} = 4\pi E/\Delta m^2_{ij}$, the oscillations do not have time to develop. The effective magnetic moment that enters into the cross-section of ES and CE$\nu$NS is then 
\begin{align}
    \mu^2_{\nu_\alpha{\rm SB}} = (\upmu^\dagger \, \upmu )_{\alpha \alpha} 
= \mu^2_{\bar{\nu}_\alpha{\rm SB}}\,.
\end{align}
Expressed in terms of the elements of the neutrino magnetic moment matrix in the mass basis, for Dirac neutrinos this effective magnetic moment takes the form   
\begin{flalign}
    & \mu^ 2_{\nu_\alpha{\rm SB}}   = |U_{\alpha1}|^2\left(|\mu_{11}|^2 + 
|\mu_{21}|^2+ |\mu_{31}|^2\right) + |U_{\alpha2}|^2\left(|\mu_{12}|^2 + 
|\mu_{22}|^2+ |\mu_{32}|^2\right)\nonumber \\ &+|U_{\alpha3}|^2
\left(|\mu_{13}|^2 + |\mu_{23}|^2+ |\mu_{33}|^2\right)+2\text{ Re}\lbrace 
U_{\alpha1}U^*_{\alpha2} \left(\mu^*_{11} \mu_{12} + \mu_{21}^* \mu_{22} + 
\mu^*_{31} \mu_{32}\right) \rbrace \nonumber \\ &+ 
2\text{ Re}\lbrace U_{\alpha1}U^*_{\alpha3} \left(\mu^*_{11} \mu_{13} + 
\mu^*_{21} \mu_{23} + \mu^*_{31} \mu_{33}\right)+ U_{\alpha2}U^*_{\alpha3} 
\left(\mu^*_{12} \mu_{13} + \mu^*_{22} \mu_{23} + \mu^*_{32} \mu_{33}\right) 
\rbrace \,,
\end{flalign}
whereas for Majorana neutrinos, 
\begin{flalign}
    \mu^ 2_{\nu_\alpha{\rm SB}} &= |U_{\alpha1}|^2\left(|\mu_{12}|^2+ 
    |\mu_{13}|^2\right) + |U_{\alpha2}|^2\left(|\mu_{12}|^2 + 
    |\mu_{23}|^2\right) +|U_{\alpha3}|^2\left(|\mu_{13}|^2 + 
    |\mu_{23}|^2\right) \nonumber \\ & +2\text{ Re}\lbrace 
    U^*_{\alpha1}U_{\alpha2} \mu_{13} \mu^*_{23}\rbrace - 2\text{ Re}\lbrace 
    U^*_{\alpha1}U_{\alpha3} \mu_{12} \mu^*_{23} \rbrace + 2\text{ Re}
    \lbrace U^*_{\alpha2}U_{\alpha3}\mu_{12} \mu^*_{13}  \rbrace \, .
\end{flalign}
Expressed through the elements of $\upmu$ in the flavour basis, the effective magnetic moment accessible in short-baseline experiments with the incoming neutrino $\nu_\alpha$ looks much simpler: 
\begin{align}
    \mu^2_{\nu_\alpha} = \sum _\beta |\mu_{\beta\alpha}|^2  \qquad \quad & 
    \text{for Dirac or Majorana neutrinos}\,. 
\label{eqn:sb_flavour_mu}
\end{align}

The existing limits reported by the experimental collaborations are summarised in Table \ref{tab:sb_limits}. As can be seen from the Table, the short-baseline accelerator experiments have constrained all three effective magnetic moments, $\mu_{\nu_e}$, $\mu_{\nu_\mu}$ and $\mu_{\nu_\tau}$, whereas reactor experiments have set upper bounds on $\mu_{\nu_e}$ using both ES and CE$\nu$NS.
\begin{table}
    \centering
    \renewcommand{\arraystretch}{1.4}
    \begin{tabular}{|c|c|c|c|}
        \hline
       Experiment  & Limit & Reference & Method   \\ \hline \hline
        LAMPF & $\mu_{\nu_e} < 1.08\times 10^{-9} \mu_B $ at 90\%C.L. & 
        \cite{Krakauer:1990cd} & Accelerator $\nu_e e^-$ \\
        LSND & $\mu_{\nu_e} < 1.1\times 10^{-9} \mu_B $ at 90\%C.L. & 
        \cite{LSND:2001akn} & Accelerator $\nu_e e^-$ \\ \hline
       
        Krasnoyarsk & $\mu_{\nu_e} < 1.4\times 10^{-10} \mu_B$ at 90\% C.L.& 
        \cite{Aleshin:2008zz} & Reactor $\bar{\nu}_e e^-$\\
        ROVNO & $\mu_{\nu_e} < 1.9\times 10^{-10} \mu_B$ at 95\% C.L.& 
        \cite{Derbin:1993wy} & Reactor $\bar{\nu}_e e^-$\\
        MUNU & $\mu_{\nu_e} < 9\times 10^{-11} \mu_B$ at 90\% C.L.& 
        \cite{MUNU:2005xnz} & Reactor $\bar{\nu}_e e^-$\\
        TEXONO & $\mu_{\nu_e} < 7.4\times 10^{-11} \mu_B$ at 90\% C.L.& 
        \cite{TEXONO:2006xds} & Reactor $\bar{\nu}_e e^-$\\
        GEMMA & $\mu_{\nu_e} < 2.9\times 10^{-11} \mu_B$ at 90\% C.L.& 
        \cite{Beda:2012zz} & Reactor $\bar{\nu}_e e^-$\\ \hline
        
CONUS & $\mu_{\nu_e} < 7.5 \times 10^{-11} \mu_B$ at 90\% C.L &  
        \cite{Bonet:2022imz}& Reactor CE$\nu$NS \\
        Dresden-II & $\mu_{\nu_e} < 2.2 \times 10^{-10} \mu_B$ at 90\% C.L & \cite{Coloma:2022avw,AtzoriCorona:2022qrf} & Reactor CE$\nu$NS \\ \hline        
        LAMPF & $\mu_{\nu_\mu} < 7.4\times 10^{-10} \mu_B$ at 90\%C.L.& 
        \cite{Krakauer:1990cd} & Accelerator $\nu_\mu e^-$ \\
        BNL-E-0734 & $\mu_{\nu_\mu} < 8.5\times 10^{-10} \mu_B$ at 90\%C.L.& 
        \cite{Ahrens:1990fp} & Accelerator $\nu_\mu e^-$ \\
        LSND & $\mu_{\nu_\mu} < 6.8 \times 10^{-10} \mu_B$ at 90\%C.L.& 
        \cite{LSND:2001akn} & Accelerator $\nu_\mu e^-$ \\ \hline
        DONUT & $\mu_{\nu_\tau} < 3.9 \times 10^{-7} \mu_B$ at 90\%C.L.& 
        \cite{DONUT:2001zvi} & Accelerator $\nu_\tau e^-$ \\ \hline
    \end{tabular}
    \caption{Current experimental constraints on the effective neutrino magnetic moments from short-baseline accelerator and reactor experiments.}
    \label{tab:sb_limits}
\end{table}

\subsubsection{\label{sec:muSolar}Effective magnetic moments for solar neutrinos}

In this work we focused on the constraints on the product of the neutrino magnetic moments and solar magnetic field strength that can be obtained from non-observation of the solar $\bar{\nu}_e$ flux. 
However, solar neutrino experiments can also constrain neutrino electromagnetic interactions through the study of the scattering of solar neutrinos on electrons. The effective magnetic moment probed in such experiments is different from the one accessible in short-baseline experiments, since in this case neutrino flavour transitions play an important role. 

We have shown that for realistic values of the solar magnetic fields and neutrino magnetic moments the flux of right-handed solar (anti)neutrinos arriving at the Earth is much smaller than that of the left-handed neutrinos; therefore, their contribution can be safely neglected when considering 
ES of solar neutrinos.%
\footnote{We have demonstrated this for Majorana neutrinos. However, from the consistency of the solar neutrino and KamLAND data, it is known that electromagnetic interactions cannot play a major role for solar neutrinos and thus, even for Dirac neutrinos, the amplitudes of the solar $\nu_R$ arriving at the Earth have to be much smaller than $\nu_L$.}
The expression for the effective magnetic moment accessible in solar neutrino experiments therefore depends only on the left-chirality amplitudes $\nu_L$ = ($\nu_{eL} \; \nu_{\mu L} \; \nu_{\tau L})^T$, which can be obtained in the standard three-flavour picture. We find 
\begin{align}
    \mu^2_{\nu{\rm SOLAR}} & = \nu^\dagger _L (\upmu^\dagger \upmu)\nu_L 
     \nonumber \\ 
    & = (|\mu_{11}|^2 + |\mu_{21}|^2 +|\mu_{31}|^2) |\nu_{1L}|^2 + 
      (|\mu_{12}|^2 + |\mu_{22}|^2 +|\mu_{32}|^2) |\nu_{2L}|^2 \nonumber \\
    & + (|\mu_{13}|^2 + |\mu_{23}|^2 +|\mu_{33}|^2) |\nu_{3L}|^2 + 
      2\rm{Re}\lbrace (\mu^*_{11}\mu_{12} + \mu^*_{21} \mu_{22} + 
      \mu^*_{31}\mu_{32}) (\nu_{1L}\nu^*_{2L})\rbrace \nonumber \\ & + 
      2 \rm{Re}\lbrace (\mu^*_{11}\mu_{13} + \mu^*_{21} \mu_{23} + 
      \mu^*_{13}\mu_{33}) (\nu_{1L}\nu^*_{3L})\rbrace \nonumber \\ 
   &+2 \rm{Re}\lbrace (\mu^*_{12}\mu_{13} + \mu^*_{22} \mu_{23} + 
     \mu^*_{32}\mu_{33}) (\nu_{2L}\nu^*_{3L})\rbrace \, .
\label{eq:munuSol1}
\end{align}

This expression is valid for both Dirac and Majorana neutrinos (it should be remembered that in the latter case the diagonal elements of the matrix $\mu$ vanish).  
 
Next, we note that the coherence of different neutrino mass eigenstates is lost on the way to the Earth, that is $\nu^*_{iL} \nu_{jL}$ averages to zero for $i \neq j$ \cite{Dighe:1999id,Grimus:2002vb}. 
Taking into account that neutrino flavour conversion in the Sun is 
adiabatic, for the probabilities of finding the mass-eigenstate components 
of the solar neutrino flux at the Earth we find 
\begin{flalign}
    |\nu_{1L}|^2 
 = c^2_{13} \cos^2\tilde{\theta}\,, 
\quad
        |\nu_{2L}|^2 = 
c^2_{13} \sin ^2 \Tilde{\theta} 
\quad \text{ and } 
    \quad     |\nu_{3L}|^2 
 = s^2_{13} \,,
\end{flalign}
where the mixing angle $\tilde{\theta}(r)$ was defined in eq.~(\ref{eq:tildetheta}) and the averaging over the coordinate of the neutrino production point in the Sun is implied.   
{}From eq.~(\ref{eq:munuSol1}) we then find 
\begin{align}
    \mu^2_{\nu{\rm SOLAR}} &= (|\mu_{11}|^2 + |\mu_{21}|^2 +|\mu_{31}|^2)
    c^2_{13} \cos ^2 \tilde{\theta} & \nonumber \\  & + (|\mu_{12}|^2 + 
    |\mu_{22}|^2 +|\mu_{32}|^2)c^2_{13} \sin ^2 \tilde{\theta} & \nonumber \\ 
    & + (|\mu_{13}|^2 + |\mu_{23}|^2 +|\mu_{33}|^2)s^2_{13} & \quad 
    \text{for Dirac neutrinos.} 
\label{eq:muDir}
\\
    \mu^2_{\nu{\rm SOLAR}} &= |\mu_{12}|^2 c^2_{13} + |\mu_{13}|^2(c^2_{13} 
    \cos^2 \Tilde{\theta} + s^2_{13}) & \nonumber \\  & + 
    |\mu_{23}|^2(c^2_{13} \sin^2 \Tilde{\theta} + s^2_{13})  & \quad 
    \text{for Majorana neutrinos.}
\label{eq:muMaj}
\end{align}
For neutrino energies $E\lesssim 1$ MeV, solar matter effects can be neglected and $\Tilde{\theta} \simeq \theta_{12}$. The effective neutrino magnetic moments can then be found from (\ref{eq:muDir}) and (\ref{eq:muMaj}) by substituting there $\tilde{\theta}=\theta_{12}$; the obtained results are in accord with eq.~(10) of \cite{Miranda:2020kwy}. For $E\gtrsim 5-7$ MeV one has $\tilde{\theta} \simeq \pi/2$. In general, to consistently extract the limits on neutrino magnetic moments from solar neutrino scattering measurements, it is important to carefully take the energy dependence of $\tilde{\theta}$ into account. 

The limits derived by Borexino and Super-Kamiokande collaborations are shown in Table \ref{tab:mu_solar}. We also include there the constraints derived in \cite{AtzoriCorona:2022jeb} from the analysis of the data of the dark matter search experiment LUX-ZEPLIN~\cite{LUX-ZEPLIN:2022qhg} as well as the recent constraints from the XENONnT experiment \cite{Aprile:2022vux}. The  excess of low-energy electron recoil events previously reported by XENON1T \cite{XENON:2020rca} is not confirmed by XENONnT and was probably due to tritium contamination \cite{Aprile:2022vux}.
\begin{table}
    \centering
    \renewcommand{\arraystretch}{1.4}
    \begin{tabular}{|c|c|c|c|}
        \hline
       Experiment  & Limit at 90\%C.L. & Reference & Energy range   
\\ \hline \hline 
Borexino & $\mu_{\nu{\rm SOLAR}} < 2.8 \times 10^{-11} \mu_B $ & 
\cite{Borexino:2017fbd,Coloma:2022umy} & 0.19 MeV $-$ 2.93 MeV \\
       \hline Super-Kamiokande & $\mu_{\nu{\rm SOLAR}} < 1.1 \times 10^{-10} 
\mu_B $  & \cite{Super-Kamiokande:2004wqk} & 5 MeV $-$ 20 MeV \\
\hline  LUX-ZEPLIN & $\mu_{\nu{\rm SOLAR}} < 6.2 \times 10^{-12} \mu_B $  & \cite{AtzoriCorona:2022jeb} &   E $\leq$ 2 MeV \\ 
\hline  XENONnT & $\mu_{\nu{\rm SOLAR}} < 6.3 \times 10^{-12} \mu_B $  & \cite{Aprile:2022vux} &  E $\leq$ 1 MeV \\ \hline
    \end{tabular}
    \caption{Limits on the effective neutrino magnetic moment from elastic scattering of solar neutrinos on electrons.} 
    \label{tab:mu_solar}
\end{table}

It should be noted that it is possible to derive stronger limits on neutrino magnetic moments than those quoted in this subsection by combining the available data on neutrino scattering, see for instance \cite{Canas:2015yoa}, where the Majorana neutrino case was considered. Such analyses can also shed some light on the so-called blind spots in the neutrino parameter space \cite{Canas:2016kfy,Sierra:2021say}.

\subsection{Other limits from astrophysics and cosmology}

\subsubsection{Plasmon decay and related processes in astrophysical environments}
Photons in plasma (plasmons) have nonzero effective mass and so can decay into neutrino-antineutrino pairs. The rate of such processes depend on effective neutrino magnetic moment given by 
\begin{equation}
    \mu^2_{\nu{\rm  PLASMON}} = \sum_{i,j} |\mu_{ij}|^2 \,.
   \label{eqn:plasmon_mu}
\end{equation}
The plasmon decay process leads to increased energy loss in stellar environments. 
By studying the impact of the extra energy loss on the luminosity of stars one can derive bounds on the neutrino magnetic moment, see Table \ref{tab:mu_astro}. In red giants, plasmon decay would be an additional source of cooling, delaying helium ignition. Non-observation of such delay was also used to constrain magnetic moments \cite{Raffelt:1992pi}. 
Besides that, additional energy losses would lead to a larger core mass at helium ignition and consequently, the tip of the red-giant branch (TRGB) would be brighter than predicted by the standard stellar models \cite{Raffelt:1999tx,Capozzi:2020cbu}.
There are also bounds on the neutrino magnetic moments from observations of the rate of change of the period of pulsating white dwarfs of spectral type DB (which have only helium absorption lines in its spectrum) \cite{C_rsico_2014}.

There are other processes contributing to stellar cooling that are sensitive to neutrino magnetic moments: for instance, $\gamma e^- \rightarrow e^- \bar{\nu}\nu$, electron-positron annihilation to neutrinos $e^+ e^- \rightarrow \bar{\nu}\nu$ and bremsstrahlung $e^- (Ze) \rightarrow (Ze)e^- \bar{\nu}\nu$. Note that all these processes probe the same combination of neutrino magnetic moments as that probed by plasmon decay. 
It has been shown that they could lead to considerable changes in the evolution of stars with masses between 7$M_\odot$ and 18 $M_\odot$, \cite{Heger:2008er}. The resulting sensitivity to the magnetic moment $\mu_{\nu{\rm PLASMON}}$ is at the level of  $(2-4) \times 10^{-11} \mu_B$. 

\begin{table}
    \centering
    \renewcommand{\arraystretch}{1.4}
    \begin{tabular}{|c|c|c|}
        \hline
       Limit & Reference & Method  \\ \hline \hline $\mu_{\nu{\rm PLASMON}} 
       < 1.2 \times 10^{-12} \mu_B $ at 95\%C.L. & \cite{Capozzi:2020cbu} & Tip of red-giant branch \\
       \hline  $\mu_{\nu{\rm PLASMON}} 
       < 1.0 \times 10^{-11} \mu_B $ at 95\%C.L. & \cite{C_rsico_2014} 
       & Pulsating white dwarfs \\
       
       \hline  $\mu_{\nu{\rm PLASMON}} < 2.2 \times 10^{-12} \mu_B $ at 
       95\%C.L. & \cite{Diaz:2019kim} & Luminosity \\ 
       \hline  $\mu_{\nu{\rm PLASMON}} < 2.2 \times 10^{-12} \mu_B $ at 
       95\%C.L. & \cite{ARCEODIAZ20151} & Luminosity \\ \hline
    \end{tabular}
    \caption{Limits on effective neutrino magnetic moments from plasmon decays in stars.} 
    \label{tab:mu_astro}
\end{table}

\subsubsection{Limits from SN1987A}
If neutrinos are Dirac particles, their nonzero magnetic moments could lead to conversion of a significant fraction of supernova (SN) neutrinos and antineutrinos into (practically) sterile $\nu_R$ and $\bar{\nu}_L$. 
For sufficiently high conversion efficiency, this would not be compatible with the observed neutrino signal from SN1987A. There are several processes that have been considered in this context and that could lead to a significant outflow of sterile neutrinos. In a hot and dense SN core, sterile neutrinos can be produced via neutrino scattering on electrons ($\nu_{L} e^ - \rightarrow \nu_Re^-$) and protons ($\nu_L p \rightarrow \nu_R p$) mediated by photon exchange, and similarly for $\bar{\nu}_R$ scattering. Once sterile neutrinos are produced, they will easily escape the SN, since their mean free path is much larger than the radius of the core. Limits based on this argument were found to be  
\cite{PhysRevLett.61.27} 
\begin{equation}
    \mu_\nu \leq (0.1-1)\times10^{-12} \mu_B\,.
\end{equation}

A detailed analysis of mediated by virtual plasmons chirality-flip neutrino scattering processes on electrons and protons in plasma was carried out in \cite{Ayala:1999xn,Kuznetsov:2009we, Kuznetsov:2009zm}. 
The following limits on flavour- and time-averaged Dirac neutrino magnetic moments were found in these papers for a number of SN models: 
\begin{equation}
    \mu_\nu < (1.1 -2.7)\times 10^{-12} \mu_B\,. 
\end{equation}
It is difficult to interpret these results in terms of more fundamental quantities since they involve weighing the contribution from different neutrino flavours depending on their abundances which vary with time.

The above limits were questioned in ref.~\cite{Bar:2019ifz}. The authors argued that a cooling proto-neutron star is not the only possible source of neutrino emission in core-collapse SN. If the canonical delayed neutrino mechanism failed to explode SN1987A, and if the pre-collapse star was rotating, an accretion disk could form. 
Neutrinos from SN1987A could have been emitted from such an accretion disk and not from the SN core. As the disc should be optically thin for neutrinos, their electromagnetic interactions would play negligible role and so would be the additional energy loss in the form of sterile neutrinos. 

\subsubsection{Conversion of $\nu_e$ from supernova neutronisation burst into $\bar{\nu}_e$}
Similarly to $\nu_e\to \bar{\nu}_e$ conversion of solar neutrinos discussed in this paper, electron neutrinos produced in SN can be  converted into electron antineutrinos due to the combined action of neutrino SFP in strong SN magnetic fields and flavour transitions  \cite{Akhmedov:1992ea,Akhmedov:2003fu,Ando:2003is,Jana:2022tsa} (note that SFP can be resonantly enhanced  in this case). 
Such a conversion would have a very clear signature for neutrinos emitted during the prompt neutronisation stage of SN evolution, as the produced neutrino flux consists almost exclusively of $\nu_e$ at this 
stage. 
The $\bar{\nu}_e$ appearance probability will depend on the product of the effective neutrino magnetic moment $\mu_\nu$ and the SN magnetic field strength $B_0$ at the resonance of SFP. 
The expression for $\mu_\nu$ takes the simplest form in a rotated (primed) basis, which differs from our primed basis defined in (\ref{eqn:basis}) by the absence of the 1--3 rotation and $\Gamma_\delta$ transformation. 
For normal neutrino mass ordering, $\mu_\nu=\mu'_{e\mu'}$, whereas for the inverse ordering $\mu_\nu=\mu'_{e\tau'}$. These quantities are related to the neutrino magnetic moments in the mass eigenstate basis as 
\begin{align}
&\mu'_{e\mu'}=\mu_{12}c_{13} e^{-i\lambda_2}+(\mu_{13}s_{12}-\mu_{23}c_{12}
e^{-i\lambda_2})s_{13} e^{i(\delta_{\rm CP} - \lambda_3)}\,,\\
&\mu'_{e\tau'}=(\mu_{13}c_{12}+\mu_{23}s_{12}e^{-i\lambda_2})e^{-i\lambda_3}\,.
\end{align}

Conversion of SN neutronisation burst $\nu_e$'s into $\bar{\nu}_e$'s can be searched for in future neutrino experiments. For example, the Hyper-Kamiokande experiment is expected to have the senstivity to $\mu_\nu B_0\sim (5\times 10^{-3}$ -- $6\times 10^{-4})\,\mu_B$\,G, depending on the neutrino mass ordering \cite{Jana:2022tsa}.
Assuming $B_0\simeq 10^{10}$\,G, this would imply the sensitivity to $\mu_\nu$  at the level of $(5\times 10^{-13}$ -- $6\times 10^{-14})$\,$\mu_B$. 

\subsubsection{Cosmology}

Neutrino magnetic moments can also be constrained by cosmology. Nonzero magnetic moments could increase the time during which neutrinos remain in thermal contact with the cosmic plasma. 
In \cite{Morgan:1981psa} the impact 
of this effect on the production of deuterium in big bang nucleosynthesis was addressed, assuming that, due to their electromagnetic scattering on electrons and positrons, neutrinos remained coupled to the plasma until the epoch of electron-positron annihilation. 

In \cite{Vassh:2015yza} the impact of transition magnetic moments of Majorana neutrinos on the neutrino decoupling temperatures and the corresponding consequences for Big Bang Nucleosynthesis were studied. Upper limits on the transition magnetic moments in the flavour basis \eqref{eqn:sb_flavour_mu} of the order $\mathcal{O}(10^{-10} \mu_B)$ were obtained.

In a different approach, a number of authors considered the production of sterile $\nu_R$ through neutrino scattering on electrons and positrons $e^{\pm}+ \nu_L \rightarrow e^{\pm} + \nu_R$ and electron-positron annihilation $e^{+} +e^{-} \rightarrow \nu_{L,R} + \bar{\nu}_{L,R}$, mediated by active-to-sterile neutrino transition magnetic moments.
Depending on the mass of the sterile neutrino states, their production can have two important consequences, (i) if sterile neutrinos are sufficiently light, they contribute to the radiation density of the Universe and modify its expansion rate, and (ii) they can also experience radiative decay $\nu_R \rightarrow \gamma+ \nu_L$, which would increase the photon energy density. Both effects can modify the primordial abundances of light elements, see for instance \cite{Dolgov:2002wy,Brdar:2020quo}.  It is difficult to interpret the results of these works in terms of fundamental magnetic moments based on the provided information on the underlying assumptions. Also, the limits have a strong dependence on the mass of the right-handed neutrino.

\subsection{Collider}
Bounds on neutrino magnetic moments are also set by collider searches for the process $e^+e^- \rightarrow \bar{\nu}\nu \gamma$ \cite{Grotch:1988ac}, including the searches for anomalous production of energetic single photons in $e^+ e^-$ annihilation at the $Z$ resonance \cite{Gould:1994gq,L3:1997exg}.  
In the latter case, the dominant mechanism for the production of single-photon events via the neutrino magnetic moment interaction is radiation of a photon from the final-state neutrino or anti-neutrino; off the resonance, it is mainly bremsstrahlung from $e^+$ or $e^-$, with the $\bar{\nu}\nu$ pair production being mediated by an $s$-channel exchange of a 
virtual photon.  
The process $e^+e^- \rightarrow \bar{\nu}\nu \gamma$  is sensitive to the same combination of the neutrino magnetic moments as the plasmon decay, eq.~\eqref{eqn:plasmon_mu}. 
The constraints coming from LEP are of the order of $10^{-6} \mu_B$ \cite{Gould:1994gq,L3:1997exg}; they are much weaker than those from astrophysical observations, but on the other hand they are more direct. 

Other processes potentially sensitive to neutrino magnetic moments could also be explored, such as e.g.\ $\pi^0 \rightarrow \gamma \bar{\nu}\nu$, but so far the obtained limits are of the same order of magnitude as those from LEP \cite{Grasso:1991qy,Grasso:1993kn}. The combination of magnetic moments constrained in these searches is the same as that in plasmon decay, eq.~\eqref{eqn:plasmon_mu}.

\section{Discussion}
\label{sec:discussion}

Assuming neutrinos to be Majorana particles, we have studied the conversion of solar $\nu_e$ into electron antineutrinos through the combined action of SFP of solar neutrinos, caused by the interaction of their transition magnetic moments with solar magnetic fields, and the ordinary flavour transitions. To this end, we have derived the neutrino evolution equations in the three-flavour framework in a rotated basis convenient for studying solar neutrinos. Making use of the fact that the effect of SFP in the Sun can at most be subleading, we developed a perturbation-theoretic approach and obtained a simple analytical expression for the probability of appearance of solar $\bar{\nu}_e$ on the Earth. The possibility that the solar magnetic fields may be twisting was taken into account. The obtained expression can be readily employed for the analysis and interpretation of the experimental results on searches of astrophysical $\bar{\nu}_e$ fluxes.

To check the validity of our approximations and the accuracy of the obtained analytical solution, we also carried out, for a number of model solar magnetic fields $B_\perp(r)$, a full numerical solution of the system of coupled neutrino evolution equations. We have found a good general agreement between our numerical and analytical results, especially for neutrino energies $E\gtrsim 5$ MeV. The discrepancy between the numerical and analytical results is larger for smaller $E$, where the $\bar{\nu}_e$ appearance probability is, however, relatively small.

We have found that the $\bar{\nu}_e$ appearance probability is to a good accuracy proportional to $[\mu_{12} B_\perp(r_0)]^2$, where $r_0$ is the coordinate of the neutrino production point in the Sun, over which averaging has to be performed. 
The contribution of the other two transition magnetic moments, $\mu_{13}$ and $\mu_{23}$, are strongly suppressed unless they exceed $\mu_{12}$ by several orders of magnitude. The shape of the profile of the solar magnetic field turns out to play relatively minor role, as the flux of the produced $\bar{\nu}_e$ is mostly determined by the average magnetic field in the neutrino production region.

With the aim to facilitate accurate analysis of and derivation of constraints from future experiments searching for solar antineutrinos, we provided the $\bar{\nu}_e$ appearance probabilities as well as the expected fluxes on the Earth in the binned form in Tables \ref{tab:estimatesAGSS09} and \ref{tab:estimatesGS98}. The calculations were done for two solar models -- low metallicity and high metallicity ones.   
We have also revisited and updated the existing upper bounds on $\mu_{12}B_\perp$ using the 3-flavour formalism developed here. The best current limit on the product of the neutrino magnetic moment and the solar magnetic field comes from the KamLAND upper bound on the astrophysical $\bar{\nu}_e$ flux, from which we have obtained $\mu_{12} B_{\perp} (r_0)\lesssim 5\times 10^{-9} \mu_B$\,kG, with a mild dependence on the solar model considered.

For reference purposes, we have also presented a comprehensive review of the other existing constrains on neutrino magnetic moments. In particular, we discussed, both for Dirac and Majorana neutrinos, how the different effective neutrino magnetic moments probed in a variety of experiments are related to the magnetic moments in the mass and flavour eigenstates bases and leptonic mixing parameters.

If the magnetic field strength in the solar core were known, one could use the upper bounds on $\mu_{12}B_\perp$ obtained from non-observation of solar $\bar{\nu}_e$ to derive constraints on $\mu_{12}$ for Majorana neutrinos. Unfortunately, very little (if anything) is known about the magnetic field strength in the core of the Sun. There is a very conservative upper bound $B< 10 ^9$\,G coming from the requirement that the pressure of the magnetic field in the solar core does not exceed matter pressure \cite{Schramm:1993mv}.
For illustrative purposes, let us assume the actual value of the magnetic field strength coincides with this upper limit.
With the KamLAND result, this would translate to the limit $\mu_{12}<5\times 10 ^{-15}\mu_B$. There are some constraints on the magnetic fields in the radiative zone of the Sun. From solar oblateness and the analysis of the splitting of the solar oscillation frequencies, one finds  $B \lesssim 7$\,MG  \cite{Friedland:2002is}.
If one assumes (rather arbitrarily) that the magnetic field in the solar core, where the neutrinos are produced, is of similar magnitude, this would translate to the limit $\mu_{12} < 7.1\times 10^{-13}\mu_B$. 
From the requirement of the stability of toroidal magnetic fields in the radiative zone of the Sun, a much more stringent limit $B\lesssim 600$\,G can be found \cite{Kitchatinov, Bonanno2013}. Assuming that the magnetic field in the core of the Sun is of similar magnitude, one would obtain the constraint $\mu_{12}<8.3 \times 10^{-9} \mu_B$. We stress once again that there is no {\it a priori} reason to believe that the magnetic fields in the core of the Sun are of the same order as those in the radiative zone; we use the latter just as some reference values.

Can one combine an independent measurement of the neutrino magnetic moment with the upper limit on $\mu B_\perp$ coming from non-observation of solar $\bar{\nu}_e$ in order to put constraints on the solar magnetic field? Assume, for example, that in the future XENONnT observes a signal that can be interpreted as being due to $\mu$-induced scattering of solar $pp$ neutrinos. Let the corresponding effective magnetic moment $\mu_{\nu\rm XENON}$, which can be obtained from eq.~(\ref{eq:muDir}) or eq.~(\ref{eq:muMaj}) by setting $\tilde{\theta}=\theta_{12}$, be about $5\times 10^{-12}\mu_B$, which is slightly below the current upper bound \cite{Aprile:2022vux}. Assuming that neutrinos are Majorana particles and that $\mu_{\nu\rm XENON}$ is dominated by $\mu_{12}$, from the discussed above KamLAND constraint on $\mu_{12}B_\perp(r_0)$ we then obtain for the magnetic field strength in the neutrino production region in the Sun  the upper limit $B_\perp< 1$\,MG.  This would apparently be the most stringent constraint on the magnetic field in the solar core currently available. However, it is obviously model dependent: it relies heavily on the assumption of significant contribution of $\mu_{12}$ to $\mu_{\nu\rm XENON}$, whereas the latter can be nonzero even if $\mu_{12}$ vanishes. It may, however, be possible to obtain a model-independent constraint on $B_\perp$ if several independent measurements of neutrino magnetic moments coming from experiments of different type become available.

The limits on the product of the neutrino magnetic moment and the solar 
magnetic field strength are expected to be improved in the near future by 
current and next-generation neutrino observatories with high potential to 
detect electron antineutrinos from astrophysical sources, which include 
Super-Kamiokande loaded with gadolinium, JUNO and Hyper-Kamiokande.
The simple analytical expression for the electron antineutrino appearance 
probability derived here as well as the calculated expected values of the 
$\bar{\nu}_e$ flux can facilitate the analyses of forthcoming data.

\acknowledgments
We thank Manfred Lindner, Mariam Tórtola and Dimitrios Papoulias for 
useful discussions.
PMM is grateful for the hospitality of the Particle and Astroparticle Physics 
Division of the \mbox{Max-Planck-Institut} f{\"u}r Kernphysik (Heidelberg) 
during the development of this project.
PMM is supported by the grants FPU18/04571 (MICIU), \mbox{PROMETEO/2018/165} 
(\mbox{Generalitat} \mbox{Valenciana}) and PID2020-113775GB-I00 
(AEI/10.13039/501100011033).

\bibliographystyle{JHEP}
\bibliography{bibliography}
\end{document}